# A Study on the Curves of Scaling Behavior of Fractal Cities

Yanguang Chen

(Department of Geography, College of Urban and Environmental Sciences, Peking University, 100871, Beijing, China. Email: chenyg@pku.edu.cn)

**Abstract:** The curves of scaling behavior is a significant concept in fractal dimension analysis of complex systems. However, the underlying rationale of this kind of curves for fractal cities is not yet clear. The aim of this paper is at researching a set of basic problems of the scaling behavior curves in urban studies by using mathematical reasoning and empirical analysis. The main findings are as follows. First, the formula of scaling behavior curves is derived from a fractal model based on hierarchical structure of urban systems. Second, the relationships between the formula of scaling behavior curves and similarity dimension are revealed. Third, according to the fractal dimension measurement methods, the scaling behavior curves are divided into two different types. Fourth, empirically, 1-dimensional spatial autocorrelation function of scaling behavior curves can be employed to reveal the basic property of scaling behavior curves. The scaling behavior curves can be utilized to evaluate fractal development extent of urban systems and identify scaling ranges of fractal cities. In positive studies, the curves may be used to help distinguish the boundaries between urban areas and rural areas and self-affine fractal structure behind the dynamical process of spatial correlation.

**Key words**: Box dimension; Cascade structure of cities; Scaling behavior curve; Fractal cities; Radial dimension; Traffic network of cities

## 1. Introduction

One of basic properties of fractals is invariance under scaling transformation. Fractals and scaling represents two different sides of the same coin. In a sense, the inherent character of fractal systems is scaling symmetry [1]. Therefore, where this are fractal phenomena, there is scaling behavior.

Fractal studies are usually associated with scaling law [2-5]. If we measure fractal dimension of cities, we always find the shadow of scaling behavior. A curve of fractal scaling behavior is a track that reflects the change of fractal dimension with measurement scale. The scaling behavior curves can be employed to identify fractal development in a city [6-9], identify urban boundaries [10], and determine scaling ranges of fractal dimension estimation. However, a series of basic problems about scaling behavior curves are still not clear for fractal and urban researchers.

This paper is devoted to exploring the mathematical essence, main types, and methods of analysis of scaling behavior curves of fractals. Using logic reasoning, mathematical derivations, numerical experiments, and positive studies, I will try to answer the following questions. First, what is the relationships and distinctions between fractal dimension and fractal scaling behavior curves? Second, how many types can fractal scaling behavior curves be divided into? Third, how to make spatial analysis of cities and regional systems by means of fractal scaling behavior curves? The rest parts will be organized as follows. In Section 2, several basic formulae will be derived from different fractal models. The mathematical essence of the scaling behavior curves will be revealed. One dimensional spatial autocorrelation function will be employed to identify the basic properties of a scaling behavior curve. In Section 3, numerical and empirical analyses based on urban traffic networks and urban growth will be made by means of theoretical models and observational data. In Sections 4 and 5, a number of related questions will be discussed, and finally, the discussion will be concluded by summarizing the main points of this study.

## 2. Models

### 2.1 Fractal dimension based on hierarchical structure

The formulae of scaling behavior curves can be derived from fractal models by means of hierarchical structure of fractal systems. Based on box-counting method, the fractal model can be expressed as an inverse power function, that is

$$N(r) = N_1 r^{-D} \to r^{-D}, \tag{1}$$

where $r$ denotes linear size of fractal copies, and $N(r)$ is the corresponding number of fractal copies, $D$ refers to fractal dimension, and $N_1$ to a proportionality constant. In theory, $N_1$=1 [11]. Thus the fractal dimension can be expressed as

$$D = -\frac{\log N(r)}{\log r}. \quad (2)$$

By differentiation of log$N$($r$) and log$r$ with respect to $r$, we have [12]

$$D = -\frac{\mathrm{d}\log N(r)}{\mathrm{d}\log r} = -\frac{\mathrm{d}N(r)/N(r)}{\mathrm{d}r/r}. \quad (3)$$

where d represents differentiation. Here equation (3) reflects a continuous fractal distribution curves. Discretizing equation (3) yields

$$D = -\frac{\Delta\log N(r)}{\Delta\log r} = -\frac{\log N(r_i) - \log N(r_{i-1})}{\log r_i - \log r_{i-1}} \doteq \frac{\Delta N(r)/N(r)}{\Delta r/r}, \quad (4)$$

where Δ refers to difference, $i$=1, 2,3,…$n$ is a rank variable, representing steps, levels of a self-similar hierarchy, and so on, and $n$ denotes the number of spatial measurement for fractal dimension.

A fractal is a hierarchy with cascade structure, which can be described with a power function or two exponential functions. This suggests that a power law can be decomposed into a pair of exponential laws based on self-similar hierarchical structure [13]. The similarity ratio for linear size of fractal copies can be defined as

$$a = \frac{r_{i-1}}{r_i}. \quad (5)$$

Accordingly, the similarity ratio for the number of fractal copies can be defined as

$$b = \frac{N_i}{N_{i-1}}. \quad (6)$$

In equations (5) and (6), $a$ and $b$ represent two similarity ratios, respectively. By recursive relations, equations (5) and (6) can be transformed into a pair of exponential functions as follows

$$r_i = \frac{1}{a} r_{i-1} = r_1 a^{1-i}, \quad (7)$$

$$N_i = bN_{i-1} = N_1 b^{i-1}. \quad (8)$$

In theory, $N_1$=1, $r_1$=1. Take the logarithm of equations (7) and (8) yields

$$\log r_i = (1-i)\log a, \quad (9)$$

$$\log N_i = (i-1)\log b. \quad (10)$$

Suppose that $i$ is a continuous variable. Derivative of equations (9) and (10) with respect to $i$ gives

$$\frac{\mathrm{d}\log r_i}{\mathrm{d}i} = -\log a, \tag{11}$$

$$\frac{\mathrm{d}\log N_i}{\mathrm{d}i} = \log b. \tag{12}$$

Combining equations (11) and (12) yields

$$D = \frac{\log b}{\log a} = -\frac{\mathrm{d}\log N_i}{\mathrm{d}\log r_i} \doteq \frac{\Delta N_i / N_i}{\Delta r / r_i}. \tag{13}$$

According to equation (2), we have

$$D = -\frac{\log N(r_i)}{\log r_i} = -\frac{\mathrm{d}\log N(r_i)}{\mathrm{d}\log r_i}. \tag{14}$$

Removing the subscript $i$ in equation (14) yields the general form of fractal dimension relation. Equation (13) is equivalent to equation (3). This suggests that fractal dimension can be evaluated by both equation (2) and equation (3). The mathematical process from equation (2) to equation (3) and then to equation (4) is simple, but the physical meaning is not clear where fractals are concerned. Equation (14) reflects the theoretical basis for defining scaling behavior curves. The curves of fractal scaling behavior in fact reflect hierarchical structure of fractal systems.

## 2.2 Similarity dimension

The formulae of scaling behavior curves are associated with similarity dimension of fractals. The similarity dimension is based on similar ratio $a$ and $b$, and can be expressed as

$$D = \frac{\log b}{\log a} = -\frac{\log(N(r_i)/N(r_{i-1}))}{\log(r_i / r_{i-1})} = -\frac{\log N(r_i) - \log N(r_{i-1})}{\log r_i - \log r_{i-1}}. \tag{15}$$

On the other hand, in light of equation (14), we have

$$D = -\frac{\log N(r_i)}{\log r_i} = -\frac{\log N(r_{i-1})}{\log r_{i-1}}. \tag{16}$$

From equation (16) it follows

$$D = -\frac{\log N(r_i) - \log N(r_{i-1})}{\log r_i - \log r_{i-1}} = -\frac{\log(N(r_{i-1})/N(r_{i-1}))}{\log(r_i / r_{i-1})} = -\frac{\Delta\log(N)}{\Delta\log(r)}, \tag{17}$$

which is equivalent to equation (15). Suppose $\Delta\log(r) \rightarrow 0$. Then we have

$$D = -\frac{\Delta\log(N)}{\Delta\log(r)} \rightarrow -\frac{\mathrm{d}\log(N)}{\mathrm{d}\log(r)}. \tag{18}$$

Apparently, equation (14) and equation (18) reach the same goal by different routes. This suggests that the similarity dimension of fractals is the just the fractal dimension of self-similarity hierarchies. The fractal scaling behavior exponent can be derived from the similarity dimension.

## 2.3 The curves of scaling behavior based on box dimension

A formula of scaling behavior curves can be given on the basis of box-counting method. For $i$=1, we have $r_0$=1, $N(r_0)$=1. In this case, equation (4) can be reduced to

$$D=\frac{\log b}{\log a}=-\frac{\log N(r_1)}{\log r_1}. \tag{19}$$

By recurrence and equal ratios theorem, we have

$$\frac{\log N(r_1)}{\log r_1}=\frac{\log N(r_2)}{\log r_2}=\cdots=\frac{\log N(r_i)}{\log r_i}. \tag{20}$$

If a system's structure deviates from the power law, equation (20) will break, and we have

$$\frac{\log N(r_1)}{\log r_1}\neq\frac{\log N(r_2)}{\log r_2}\neq\cdots\neq\frac{\log N(r_i)}{\log r_i}. \tag{21}$$

This suggests that non-fractals, multifractals, or scale-dependence fractals cannot satisfy equation (20). For multifractals or scale-dependence fractals, equation (3) can be generalized to a function such as

$$D(r)=-\frac{\mathrm{d}\log N(r)}{\mathrm{d}\log r}, \tag{22}$$

which $D(r)$ represents the extended fractal dimension. This is treated as the scale-dependence fractal dimension [12].

In fact, scale-dependence fractal dimension may suggest multifractal scaling or self-affine fractal scaling. Standard fractal structure indicates isotropic growing fractals, while self-affine scaling indicates anisotropic growing fractals. Equation (22) can be discretized as

$$D^*(r)=-\frac{\Delta\log N(r)}{\Delta\log r}=-\frac{\log N(r_i)-\log N(r_{i-1})}{\log r_i-\log r_{i-1}}, \tag{23}$$

which is similar to but different from equation (15). For simplicity, equation (23) can be re-expressed as a special formula of scaling behavior curve as follows

$$\alpha_i = -\frac{\log N_i - \log N_{i-1}}{\log r_i - \log r_{i-1}} \to D, \tag{24}$$

in which $N_i=N(r_i)$, $N_{i-1}=N(r_{i-1})$, $\alpha_i=D^*(r)$. Here $a_i$ represents the scaling behavior exponent based on inverse power law. This is the formula of scaling behavior curve for box dimension.

## 2.4 The curve of scaling behavior based on radial dimension

To describe scale-free phenomena of growing fractals based on core-periphery relationship, we need number-radius scaling. The number-radius scaling can be replaced by area-radius scaling. This method gives a local fractal parameter termed *radial dimension* [14]. In theory, a radial dimension may be equal to a box dimension for a regular growing fractal [6]. However, for a random fractal, due to method of spatial measurement or data extraction, a radial dimension is often different in value from its corresponding box dimension. Moreover, a box dimension value is independent of growing center, while a radial dimension value depends on the definition or selection of growing center. In this sense, box dimension represents a global fractal parameter, while radial dimension represents a local fractal parameter [7]. A box dimension is usually given by an inverse power law, while a radial dimension is always given by a positive power law. The model for the radial dimension is as follows

$$N(r) = N_1 r^D \to r^D, \tag{25}$$

where $r$ is the distance from the growth center, $N_1$ denotes proportionality coefficient, and $D$ refers to radial dimension. Where area- or number-radius scaling method is concerned, $r$ denotes the radii of concentric circles. Specially, we have $N_1=1$. The fractal dimension can be expressed as

$$D = \frac{\log(N(r)/N_1)}{\log r}. \tag{26}$$

If the core-periphery relation deviates from the power law, we have

$$D(r) = \frac{\mathrm{d}\log(N(r)/N_1)}{\mathrm{d}\log r} = \frac{\mathrm{d}\log(N(r))}{\mathrm{d}\log r} = \frac{\mathrm{d}N(r)/N(r)}{\mathrm{d}r/r}, \tag{27}$$

which can be discretized as

$$D^*(r) = \frac{\Delta\log(N(r))}{\Delta\log r} \doteq \frac{\Delta N(r)/N(r)}{\Delta r/r}. \tag{28}$$

The curve of fractal scaling behavior can be described by

$$\alpha(r_i)=\frac{\log(N(r_i))-\log(N(r_{i-1}))}{\log r_i-\log r_{i-1}}\doteq\frac{(N(r_i)-N(r_{i-1}))/N(r_{i-1})}{(r_i-r_{i-1})/r_{i-1}},\tag{29}$$

where $a_i$ represents the scale-dependence scaling exponent. According to custom, the symbol $D^*(r)$ is replaced by $\alpha(r_i)$. For simplicity, equation (29) can be expressed as [7]

$$\alpha_i=\frac{\log N_i-\log N_{i-1}}{\log r_i-\log r_{i-1}}.\tag{30}$$

This is the formula for the scaling behavior curves based on radial dimension, which is in contrast to equation (24) based on box dimension. The parameter $a_i$ represents the scaling exponent based on power law indicating fractal structure.

## 2.5 Autocorrelation function based on scaling behavior curve

The key of scaling behavior lies in scaling relation of a system. One of basic properties of fractals is scaling invariance, which indicates dilation symmetry. The scaling law of fractal systems can be expressed as following scaling relation

$$N(\zeta r)=\zeta^{\pm D}N(r),\tag{31}$$

where $\zeta$ denotes a ratio of scaling up or down. For scale-free scaling behavior curve, we have

$$\frac{\mathrm{d}\log N(\zeta r)}{\mathrm{d}\log(\zeta r)}=\frac{\mathrm{d}(\log(\zeta^{\pm D})+\log N(r))}{\mathrm{d}(\log(\zeta)+\log(r))}=\frac{\mathrm{d}\log N(r)}{\mathrm{d}\log(r)}.\tag{32}$$

This suggests that, for a fractal, the scaling behavior curve follows scaling law. If the scaling behavior curve departs from scaling relation, we should find the reason and effect.

For a regular fractal, a scaling behavior curve is actually a horizontal straight line rather than a curve. The scaling exponent is equal to fractal dimension, that is, $a_i \equiv D$. In contrast, for a random fractal, a scaling behavior curve is the curve that changes randomly around a horizontal straight line. The value of scaling exponent changes around the corresponding fractal dimension. Therefore, the 1-dimension spatial autocorrelation function (ACF) and the partial autocorrelation function (PACF) can be employed to evaluate the development level of fractal systems. Autocorrelation analysis come from time series analysis theory and method [15], but it can be directly generalized to 1-dimension spatial series analysis [16]. Based on the formula for scaling behavior curve, the 1-dimensional ACF can be defined as follows

$$\hat{\rho}(k)=\frac{\sum_{i=k+1}^{n-2}[(\alpha_i-\bar{\alpha})(\alpha_{i-k}-\bar{\alpha})]}{\sum_{i=1}^{n-2}(\alpha_i-\bar{\alpha})^2}. \tag{33}$$

The PACF can be calculated by Yule-Walker's recurrence formula [15]. Correspondingly, the standard error can be approximately estimated by:

$$S=\frac{1}{\sqrt{n-1}}. \tag{34}$$

Note that the number of points of scaling behavior curve is $n$-1 instead of $n$. Based on the ACF and PACF, histograms and what is called "two-standard-error bands", can be used to show whether or not there is significant difference between zero and ACF or PACF values [15]. What is called "two-standard-error bands" consists of twice the positive standard errors and twice the negative standard errors. The two times of positive and negative standard errors form two transversal curves (For ACF) or lines (For PACF) on a correlogram.

# 3. Empirical analysis

## 3.1 Mathematical experiment for behavior curves

The random fractals in the real world are different from the regular fractals in the mathematical world. Regular fractals have clear self-similarity, while random fractals exhibit statistical self-similarity that is not easily perceived intuitively. In addition, there is a difference between self-similar fractals and self-affine fractals. The intuitive auxiliary tools for identifying statistical self-similarity or self-affinity are various coordinate graphs or curve charts. The first intuitive tool is the log-log plot, which is used to reflect the power-law relationship between the measurement scale and the corresponding measure. The second intuitive tool is the scaling behavior curve. This is the focus of this study. The third intuitive tool, as indicated above, is the correlogram (ACF correlogram) of the behavior curve. For the convenience of readers to read the case analyses in this paper and conduct fractal research in future, it may be helpful to tabulate the key points for examining illustrations as follows (Table 1).

**Table 1 Key points for reading figures including log-log plot, scaling/non-scaling behavior curve chart and correlogram**

| Graph | Type | Scatter point distribution | Fractal or not |
|---|---|---|---|
| **Log-log plot** | Type 1 | Sloped straight line trend | Well developed fractals |
| | Type 2 | Curve with sufficiently long sloped straight line segments in the front or middle as scaling range | Fractals with restricted development |
| | Type 3 | Curve consists of two inclined straight line segments with different slopes | Bi-fractals reflecting self-affine growth |
| | Type 4 | All or most of points cannot match a straight trend line | Non fractal structures, or fractal structures that have not yet emerged |
| **Behavior curve chart** | Type 1 | All points randomly fluctuate around a horizontal straight line | Well developed fractals |
| | Type 2 | Random fluctuation with one horizontal straight line trend segment | Fractals with restricted development |
| | Type 3 | Random fluctuation with two horizontal straight line trend segments | Bi-fractals reflecting self-affine growth |
| | Type 4 | No obvious horizontal trend segment | The situation is diverse, needing further research |
| **Correlogram (ACF histogram)** | Type 1 | No autocorrelation coefficient exceed “two-standard-error bands" | No significant autocorrelation |
| | Type 2 | A few autocorrelation coefficient exceed “two-standard-error bands" | Weak significant autocorrelation |
| | Type 3 | A number of autocorrelation coefficient exceed “two-standard-error bands" | Strong significant autocorrelation |

**Note**: Type 1 is a relatively ideal situation, which is rare in reality. Type 2 is the most common situation in the real world.

Not all behavior curves are scaling behavior curve. If and only if a behavior curve is associated with scaling relation, it can be treated as scaling behavior curve. A simple mathematical experiment can be utilized to draw a comparison between scaling behavior curve and non-scaling behavior curve. Two mathematical models for cities can be employed to illustrate the differences between two types of behavior curves: scaling behavior curve and non-scaling behavior curve (Figures 1, 2, 3). One is Clark’s model for urban population density distribution [17], and the other is Smeed’s model for traffic network density distribution of cities [18]. Clark’s model reflects the geographical phenomena with characteristic scales and can be described with conventional mathematical methods [12], while Smeed’s model reflects the geographical phenomena without characteristic scale and cannot be described with conventional mathematical method [6]. In short, Clark’s exponential decay process bears an effective mean representing characteristic length, while Smeed's inverse law decay bear no determined mean and thus bear no characteristic length. The average value of Smeed's distribution depends on the scales of measurement and must be analyzed with the ideas from scaling symmetry. Therefore, Clark's model can be used to generate a non-scaling behavior curve, while Smeed's model can be used to produce a scaling behavior curve. Due to the connection of the scaling exponent of Smeed's model with fractal dimension, we can create a standard fractal scaling behavior curve by using this inverse power function (Supplementary Data S1).

First of all, let's examine the non-scaling behavior curve based on negative exponential decay. Clark's model is suitable for describing the density decay regularity of urban population distribution [19, 20]. The standard form of Clark's model is

$$\rho(r)=\rho_0 e^{-r/r_0}, \tag{35}$$

where $r$ is the distance to city center, $\rho(r)$ is the average population density corresponding the radius $r$, the coefficient $\rho_0$ represents the population density of city center, and $r_0$ refers to the characteristic radius of population distribution. The parameter $r_0$ represents characteristic length of the spatial distribution of urban population density. For a density distribution, integrating $\rho(r)$ over $r$ based on 2-dimension space yields a cumulative distribution, that is

$$P(r)=2\pi\int_0^r x\rho(x)\mathrm{d}x, \tag{36}$$

where $P(r)$ denotes the total population within the scope of radius $r$. Where Clark's model is concerned, the cumulative distribution function is

$$P(r)=2\pi\rho_0\int_0^r xe^{-x/r_0}\mathrm{d}x=2\pi r_0^2\rho_0[1-(1+\frac{r}{r_0})e^{-r/r_0}]. \tag{37}$$

Suppose that $\rho_0$=30000, $r_0$=5. We have

$$P(r)=1500000\pi[1-(1+\frac{r}{5})e^{-r/5}]. \tag{38}$$

The scaling behavior can be expressed as

$$\alpha_i=\frac{\log P_i-\log P_{i-1}}{\log r_i-\log r_{i-1}}. \tag{39}$$

In theory, equation (39) can be approximated by the following formula

$$\alpha_i=\frac{(P_i-P_{i-1})/P_{i-1}}{(r_i-r_{i-1})/r_{i-1}}. \tag{40}$$

As far as equation (40) is concerned, the precondition of effective calculation is that $r$ is a continuous variable.

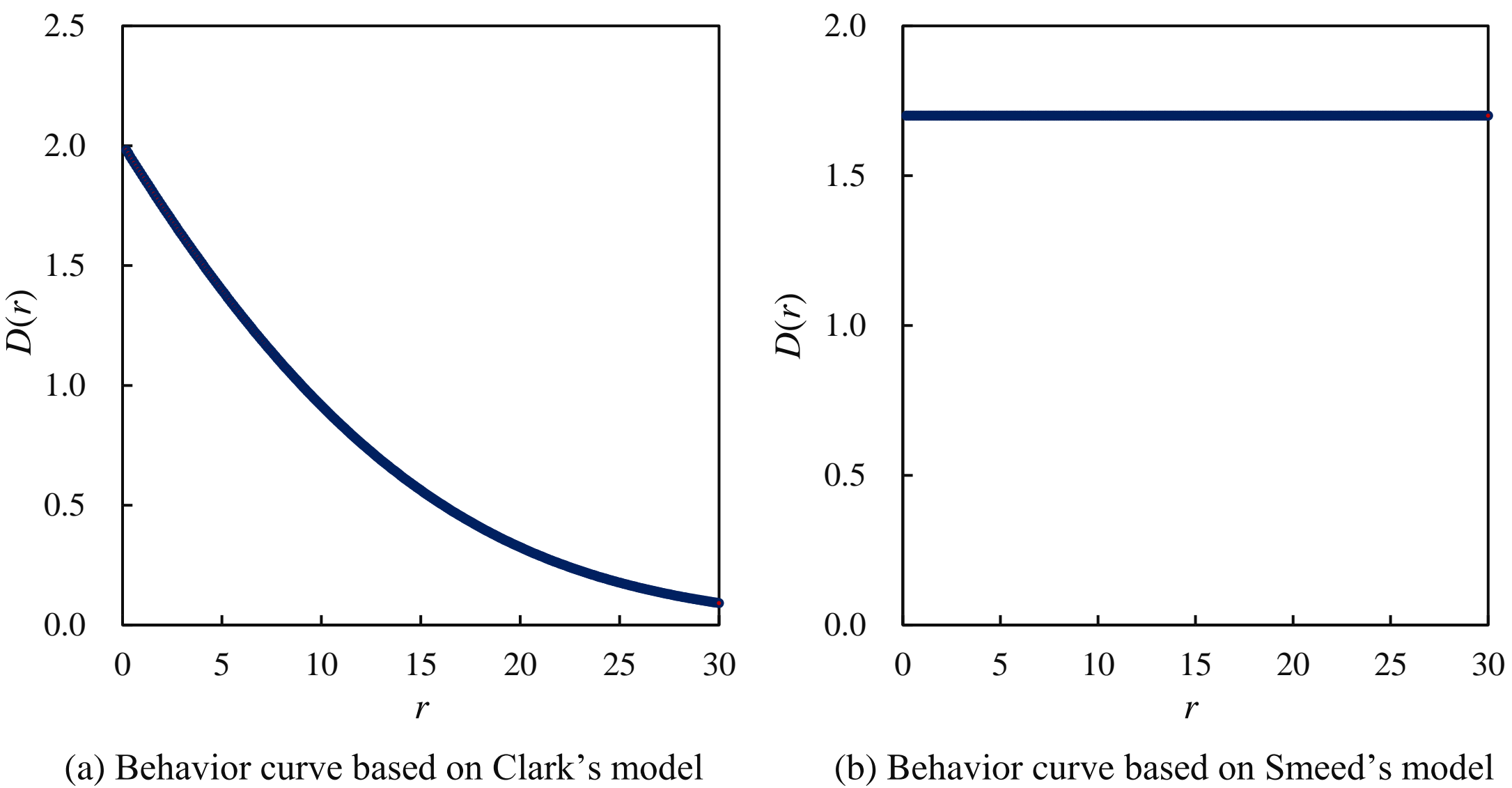

(a) Behavior curve based on Clark's model (b) Behavior curve based on Smeed's model

**Figure 1 The standard behavior curves of the urban density based on Clark's model and Smeed's model**

(**Note**: The first behavior curve, Figure 1(a), is generated by using equations (38) and (39), the second behavior curve, Figure 1(b), is created by using equations (30) and (43). The former is not a scaling behavior curve, while the latter is a regular fractal scaling behavior curve.)

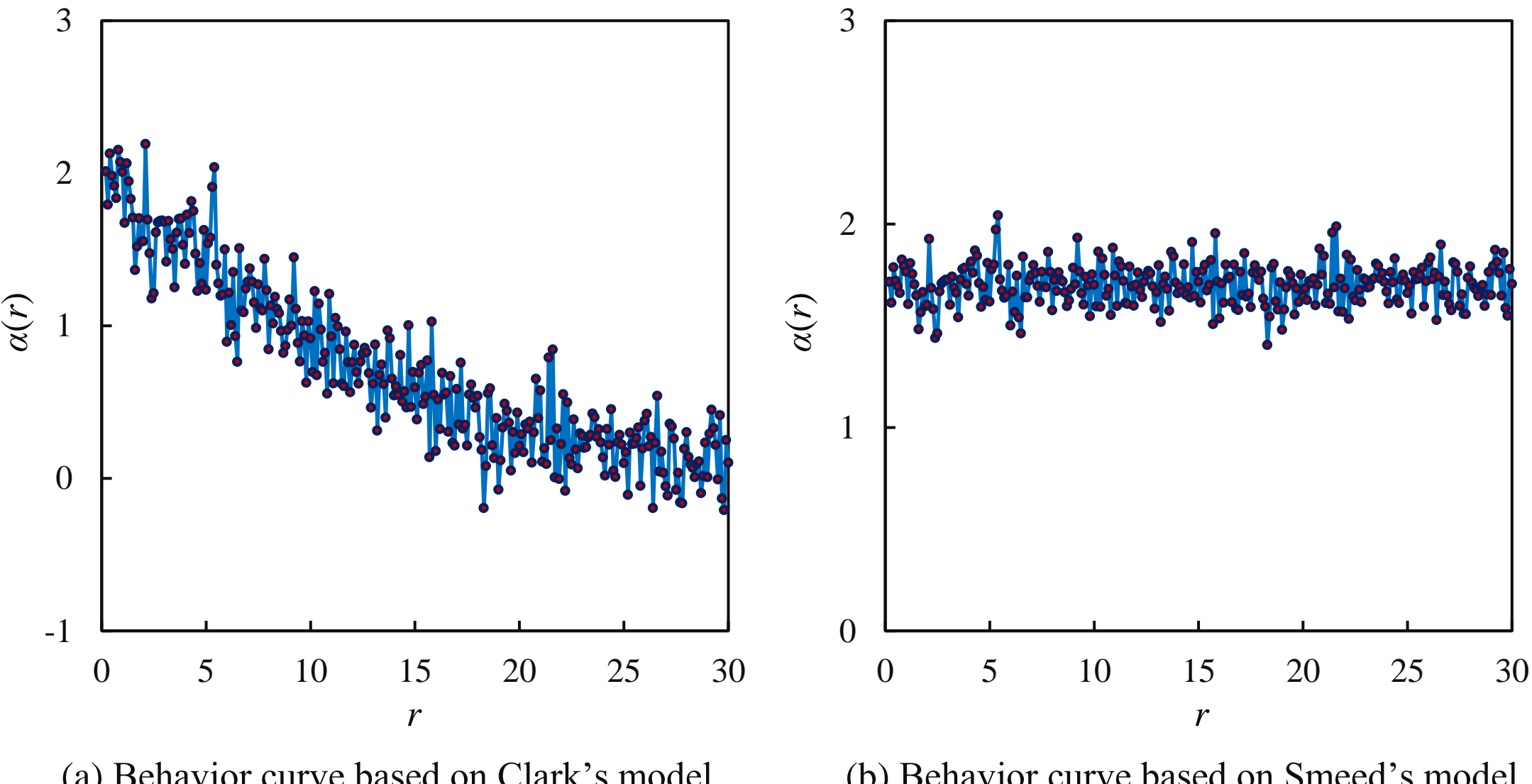

(a) Behavior curve based on Clark's model (b) Behavior curve based on Smeed's model

**Figure 2 The random behavior curves of the urban density based on Clark's model and Smeed's model**

(**Note**: The first behavior curve, Figure 2(a), is generated by introducing normal white noises into the curve in Figure 1(a), the second behavior curve, Figure 2(b), is created by introducing white noises into the curve in Figure 1(b). The former is not a scaling behavior curve, while the latter is a random fractal scaling behavior curve.)

Then, let's investigate the behavior curve based on inverse power law decay. Smeed's model is suitable for describing the density decay regularity of urban traffic [21]. The standard form of Smeed model is

$$\rho(r)=\rho_1 r^{D-d}, \tag{41}$$

where $\rho_1$ represents the traffic network density near city center, $D$ is the fractal dimension of traffic network, and $d$=2 refers to Euclidean dimension of embedding space [6, 22]. The other symbols are the same as those in equation (35). In terms of equation (36), the cumulative distribution of traffic network is

$$N(r)=2\pi\rho_1\int_0^r x^{D-d+1}\,\mathrm{d}\,x=\frac{2\pi\rho_1}{D}r^D, \tag{42}$$

where $N(r)$ denotes the total length of streets and roads or total number of traffic nodes within the scope of radius $r$. Suppose that $\rho_1$=10000, $D$=1.7. We have density distribution $\rho(r)$=10000$r^{-0.3}$. So, the cumulative distribution is

$$N(r)=\frac{20000\pi}{1.7}r^{1.7}. \tag{43}$$

Thus, the scaling behavior curve can be expressed as equation (30), which can be approximated by

$$\alpha_i=\frac{(N_i-N_{i-1})/N_{i-1}}{(r_i-r_{i-1})/r_{i-1}}. \tag{44}$$

Where equation (44) is concerned, the precondition of effective calculation is also that $r$ is a continuous variable.

**Table 2 A comparison between non-scaling behavior curve and scaling behavior curve**

| **Type** | **Non-scaling behavior curve based on Clark's model** | **Scaling behavior curve based on Smeed's model** |
|---|---|---|
| **Standard curve** | An attenuation curve | Horizontal straight line |
| **Actual curve** | An attenuation curve attached white noise | A horizontal straight line attached white noise |
| **ACF** | A number of autocorrelation coefficients exceed "two-standard-error bands" | No (in theory) or a few (in practice) autocorrelation coefficients exceed "two-standard-error bands" |
| **Property** | With characteristic scale (not following the scaling law) | Scaling invariance (following the scaling law) |

**Note**: The scaling behavior curve of a real-world fractal is often not a horizontal straight line, but the local trend line take on a horizontal line. Generally speaking, the middle section shows a horizontal straight trend, and the straight segment is regarded as *scaling range*.

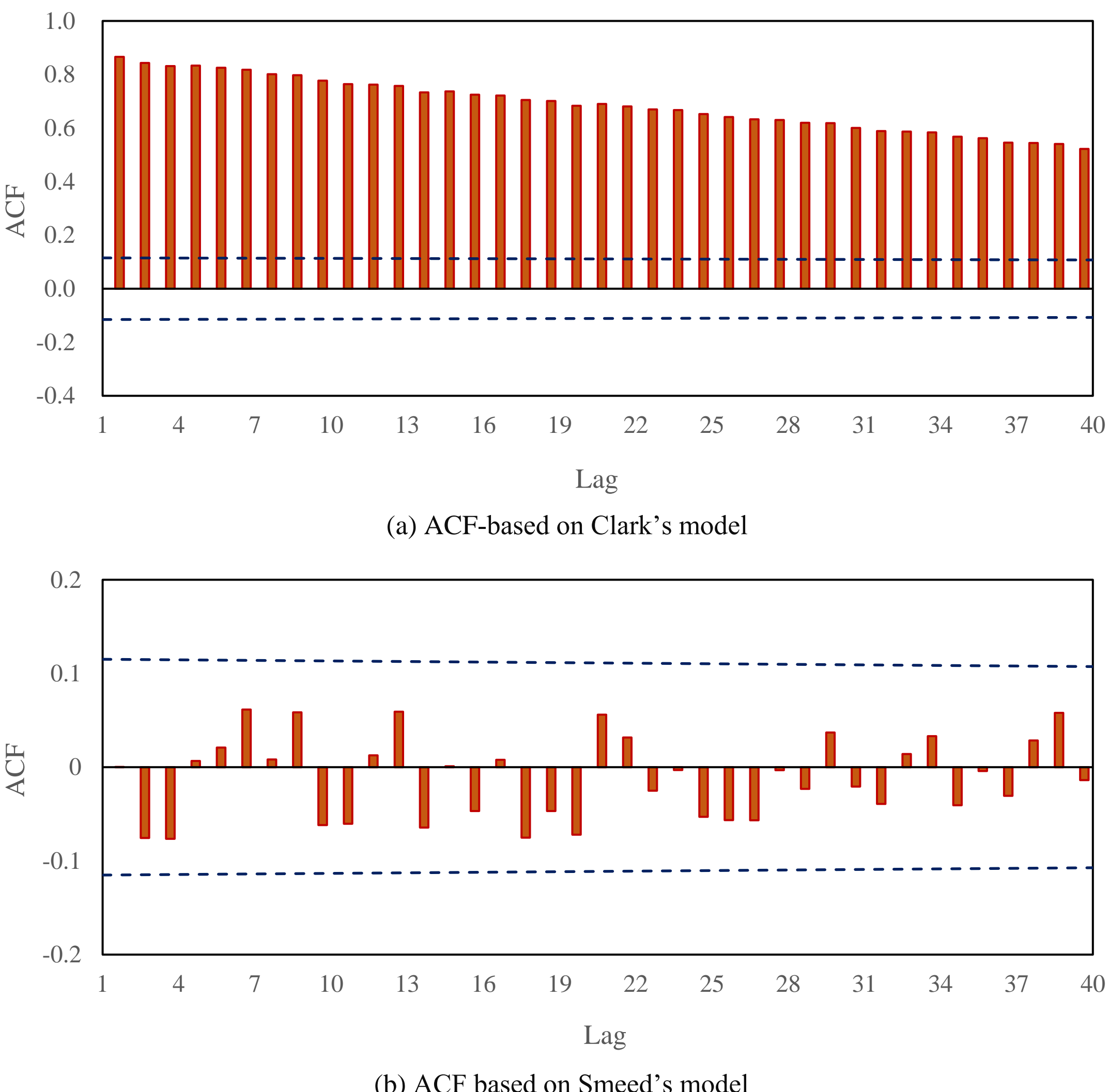

(a) ACF-based on Clark's model

(b) ACF based on Smeed's model

**Figure 3 Histograms of 1-dimensional spatial ACF of the non-scaling behavior curve based on Clark's model and scaling behavior curves based on Smeed's model**

(**Note**: It is easy to calculate ACF and PACF using IBM SPSS, the path is as follows: Analyze→ Forecasting → Autocorrelations. The ACF values are computed by using equation (33). The two dashed lines in the histograms is called "two-standard-error bands", according to which we can know whether or not there is significant difference between zero and ACF or PACF values. See Supplementary Data S1 for the PACF.)

If behavior curves are generated by mathematical models, the results will be in sharp contrast. The differences help us distinguish fractal scaling behavior curve from the non-scaling behavior curves in the real world. Using equations (38) and (39), we can generate a standard non-scaling behavior curve based on Clark's model (Figure 1(a)). Introducing a white noise sequence following normal distribution into the curve yields a random non-scaling behavior curve (Figure 2(a)). Based on the random non-scaling behavior curve, we can calculate the ACF values (Figure 3(a)). In contrast, using equations (43) and (44), we can create a standard scaling behavior curve based on Smeed's model for fractal cities (Figure 1(b)). Introducing a normal white noise sequence into this curve yields a random scaling behavior curve (Figure 2(b)). Based on the random non-scaling behavior curve, we can calculate the ACF values (Figure 3(b)). The difference between the two types of behavior curve can be tabulated as follows (Table 2).

### 3.2 Case analysis of Changchun city

In the real world, scaling behavior is always associated with complex systems. If a system bear a single scaling process, the scaling behavior curve takes on a horizontal straight line. However, if a system possesses multiple scaling processes, which are mixed together, the scaling behavior curve will be more complicated than the one based on standard model and it is hard to differentiate it from a non-scaling behavior curve. By means of ACF, we can reveal some fractal property of complex systems through fractal scaling behavior curves. One of complex geographical systems is traffic or transportation network, which proved to bear fractal nature [22-26]. A number of studies show that urban road and street networks can be described with fractal geometry [27-33]. Moreover, railway networks can also be treated as fractal systems [34, 35]. In this case, the fractal scaling behavior curve can be used to study the fractal characteristics of traffic network. An example is to find the fractal growing center for the traffic network of Changchun city of China by using scaling behavior curves (Supplementary Data S2). Two measurements can be utilized to describe traffic fractals: one is traffic lines, i.e., streets and roads, and the other is points, i.e., nodes of traffic lines. There are two possible fractal growing centers for Changchun city's traffic networks: one the railway station, and the other, central business district (CBD) [22].

The traffic network of Changchun city takes on statistically self-similar properties. Both the intra-urban traffic lines and traffic nodes follows fractal scaling law. Smeed's model can be well fitted the observational data of urban density distribution based on road lengths and node numbers. In comparison, traffic lines seem to exhibit more significant fractal properties than traffic nodes. On the other, the fractal growth centered on railway station seems to be more significant than that centered on CBD (Figure 4). The fractal scaling behavior curve based on traffic lines seem to be more stable than ones based on traffic nodes (Figure 5). However, the curve in Figure 5(a) is difficult to distinguish from the curve in Figure 5(b), and the curve in Figure 5(c) is also difficult to distinguish from the curve in Figure 5(d). In this case, we need the help of an ACF histogram.

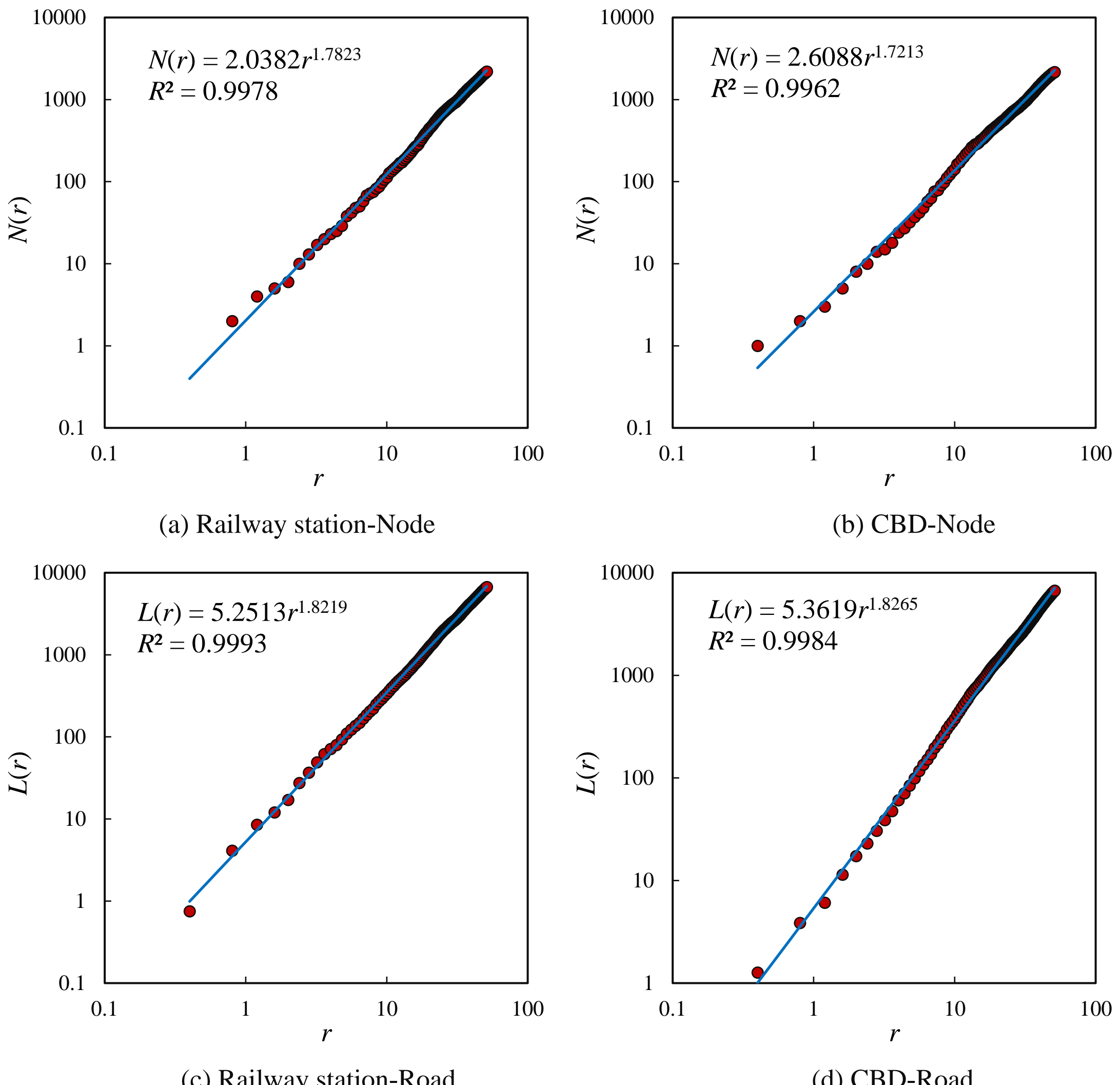

**Figure 4 The double logarithmic relationships between the distance to city centers and total length of road lines of Changchun city, China (2006)**

(**Note**: The original data were cited from the complementary materials of Chen *et al* [22]. See Supplementary Data S2 for the processed results.)

For Changchun city, railway station represents the starting point of city development, and CBD reflects the current possible center of the city. ACF provides important information for evaluating the traffic fractals. The results are as below. The ACF values of the scaling behavior curves of traffic nodes and lines centered on railway station shows no significant autocorrelation. This is close to standard random scaling behavior curve and suggests a believable fractal structure (Figures 6(a) and (c)). The ACF values of the scaling behavior curve of traffic lines and nodes centered on CBD displays weak autocorrelation (Figure 6(b) and (d)). The self-similar fractal structure is not very satisfying. A conclusion can be reached that, as far as fractal growing is concerned, the city center of Changchun is railway station rather than CBD. This inference is helpful for us to know the city development from the prospective of urban evolution.

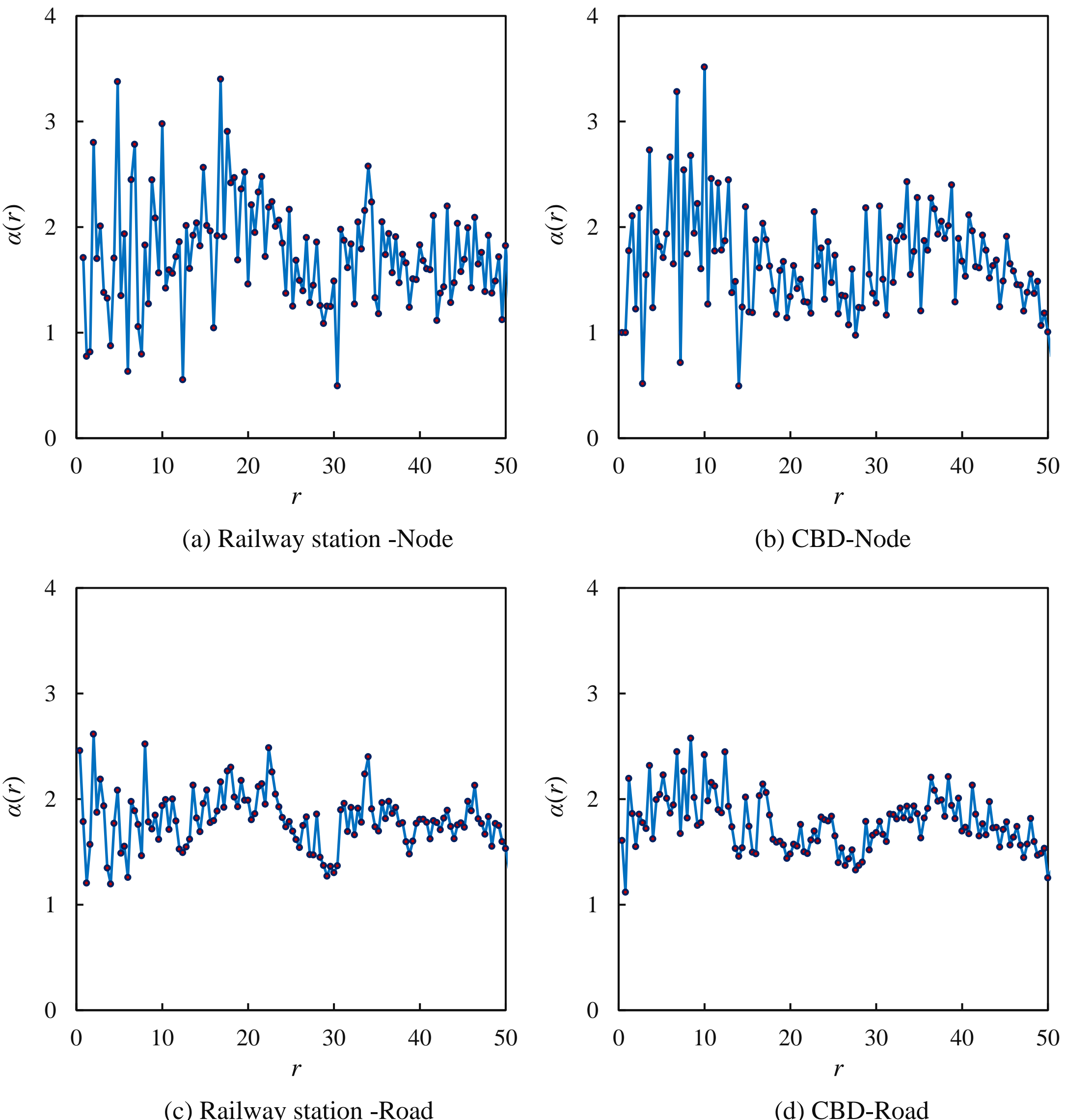

**Figure 5 The fractal scaling behavior curves of the traffic network of Changchun city, China (2006)**

(**Note**: The scaling behavior curves are given by using equation (30), but for traffic lines, the number $N(r)$ was replaced by total length $L(r)$. The scaling behavior curves based on railway stations are very similar to the corresponding scaling behavior curves based on CBD. correlograms can be used to obtain feature information.)

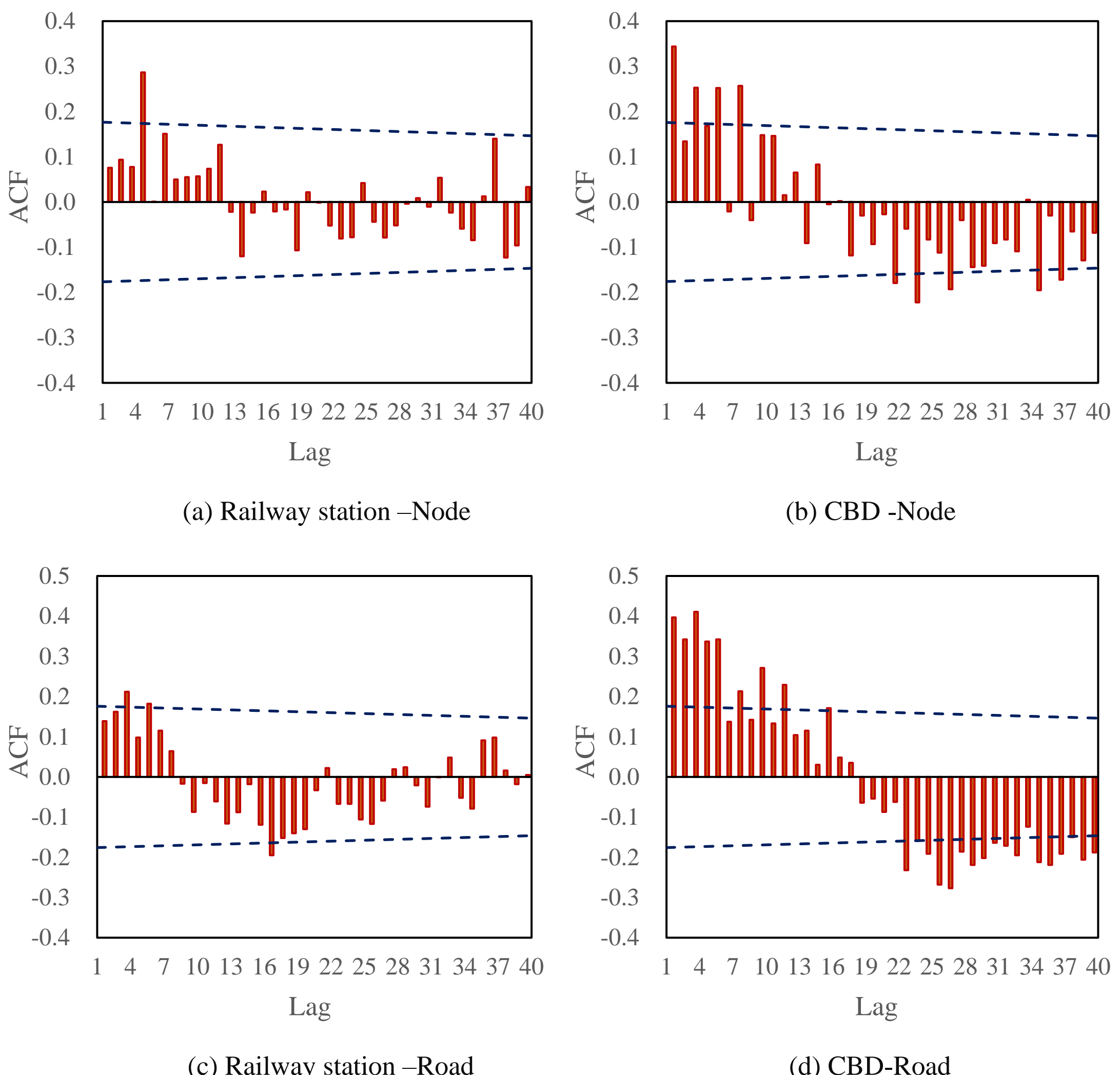

(a) Railway station –Node

(b) CBD -Node

(c) Railway station –Road

(d) CBD-Road

**Figure 6 Histograms of 1-dimensional spatial ACF of the scaling behavior curves of Changchun's traffic network**

(**Note**: The two lines in the histograms is called "two-standard-error bands", according to which we can know whether or not there is significant difference between zero and ACF values. See attached Supplementary Data S2 for PACF. Compared with Figure 6(a), there are more autocorrelation coefficient values in Figure 6(b) that exceed two-standard-error bands; compared with Figure 6(c), there are more autocorrelation coefficient values in Figure 6(d) that go beyond the two-standard-error bands.)

## 3.3 Case analysis of Beijing and Guangzhou cities

As a comparison, the next study is Beijing traffic network and Guangzhou traffic network. The property and source of original data of the two cities for traffic systems differ from that of Changchun city [21] (Supplementary Data S3). In light of log-log plot reflecting the scaling relationships between urban radius and traffic road length, Beijing's traffic network departs significantly from the scaling trend line, while Guangzhou's traffic network takes on scaling character to a degree (Figure 7). The scaling behavior curves deviates from the horizontal trend line to some extent. For Guangzhou's traffic network is concerned, the departure is not very serious. However, for Beijing's traffic network, the scaling behavior curve looks like the scaling behavior curve based on Clark's model instead of that based on Smeed's model (Figure 8). Actually, the traffic network density can be modeled by a gamma function rather than an inverse power function. The ACF values of the scaling behavior curve of Beijing's traffic network show strong autocorrelation. Relatively, autocorrelation of the scaling behavior curve of Guangzhou's traffic network is not so strong (Figure 9).

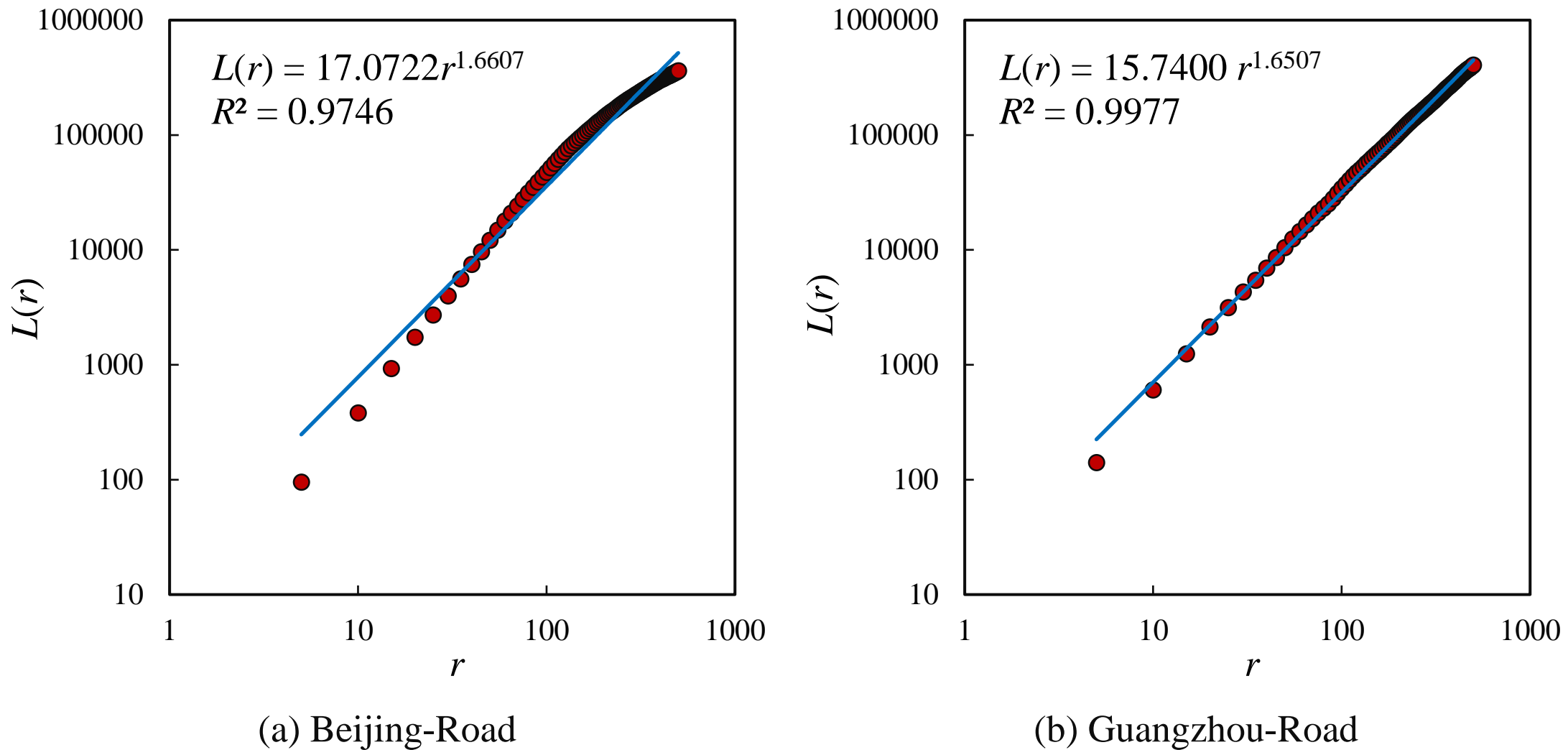

(a) Beijing-Road (b) Guangzhou-Road

**Figure 7 The double logarithmic relationships between the distance to city centers and total length of road lines of Beijing and Guangzhou cities, China (2016)**

(**Note**: The original data were cited from the complementary material of Chen and Long [21]. For Beijing, the trend line based on power function cannot be well fitted to the scatter points. This suggests that Beijing's traffic network does not follow distance-decay scaling law. Guangzhou's traffic network can be approximately modeled with inverse power law, and the matching effect between power-law trend lines and scatter points is better. )

If a scaling behavior curve takes on a horizontal trend line with random fluctuations, it suggests fractal property. On the other hand, if the scaling behavior curve depart horizontal trend line, the thing cannot be judged simply. For a self-similar fractal city of isotropic growth, the scaling behavior curve shows a horizontal trend line. However, for a self-affine fractal city of anisotropic growth, the density distribution curve looks similar to negative exponential decay, indicated by Clark's model. In this case, the fractal scaling behavior curve may be an exponential-like curve rather than a horizontal straight line. If the geographical phenomenon is a multifractal system, the scaling behavior may be very complicated. A fractal city in the real world is always the result of the superposition of multiple fractal structures.

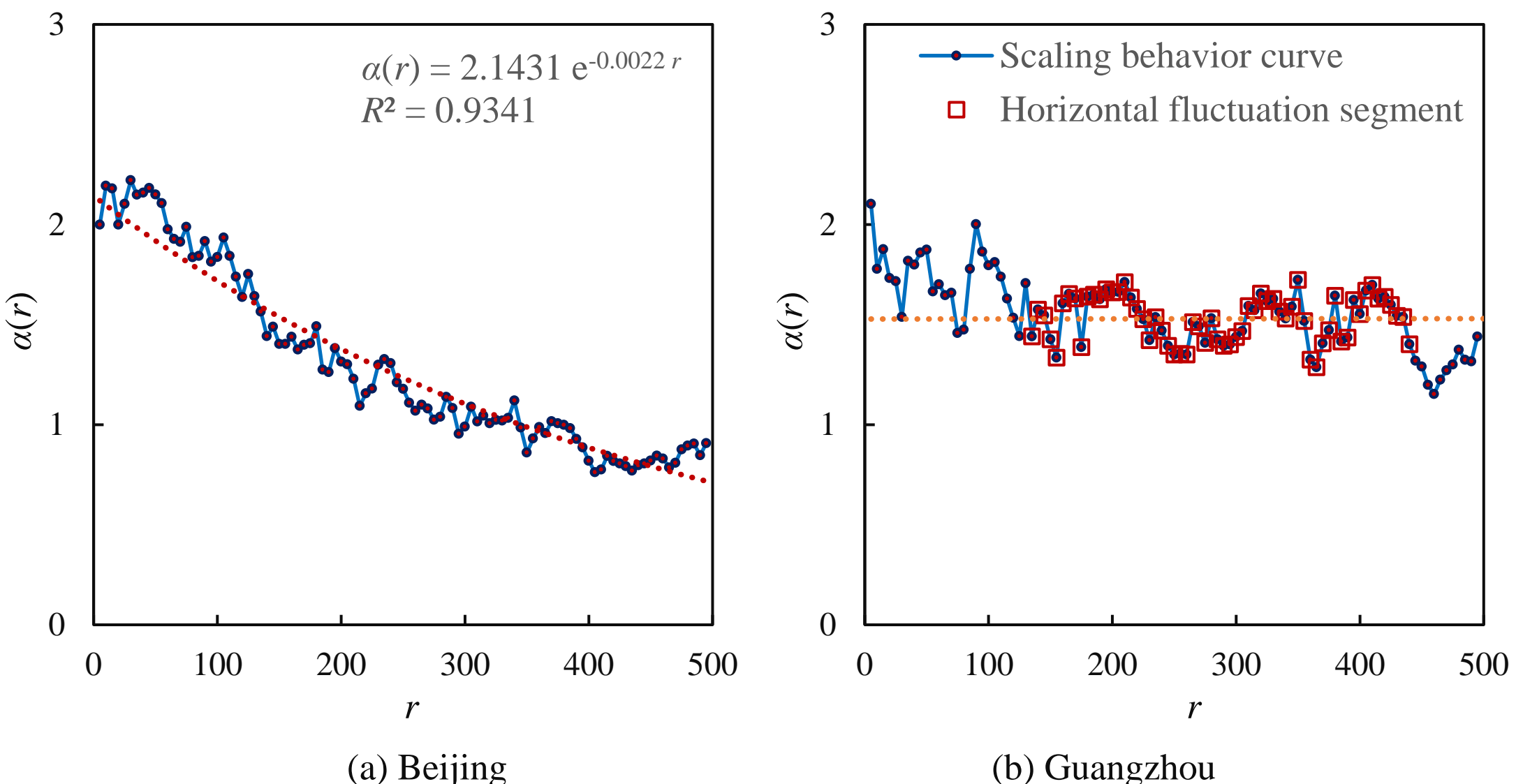

(a) Beijing (b) Guangzhou

**Figure 8 The fractal scaling/non-scaling behavior curves of the road networks of Beijing and Guangzhou cities, China (2016)**

(**Note**: The scaling behavior curves are created by using equation (30), but the number $N(r)$ was replaced by total length $L(r)$. The scaling behavior curve of the traffic networks in Beijing shows a significant exponential decay trend, while the scaling behavior curve of Guangzhou has a horizontal fluctuation segment in the middle part. The difference between the two curves can be further reflected by the autocorrelation functions.)

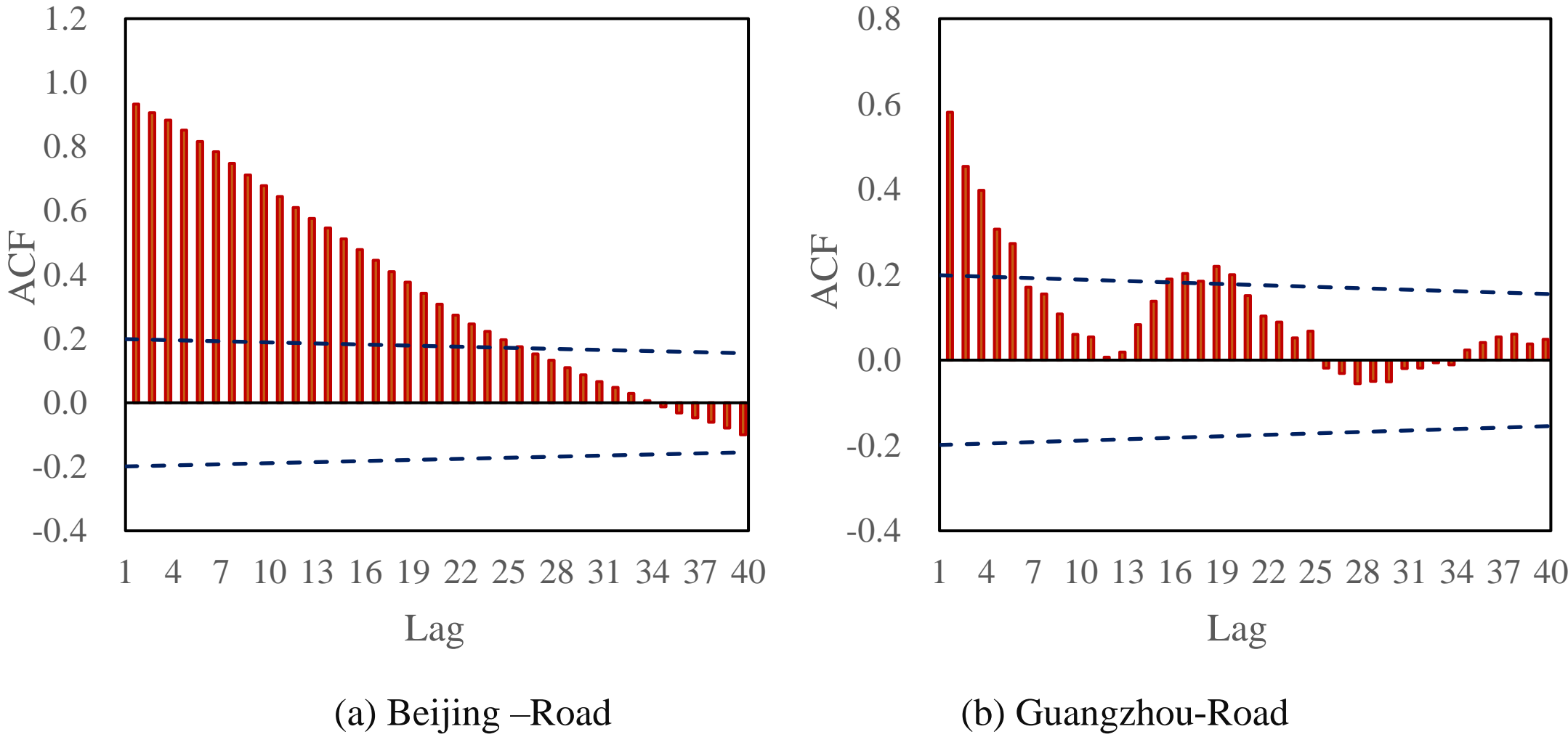

(a) Beijing –Road (b) Guangzhou-Road

**Figure 9 Histograms of 1-dimensional spatial ACF of the scaling/non-scaling behavior curves of Beijing's and Guangzhou's traffic networks**

(**Note**: For standard fractal structures, the scaling behavior curve does not show significant autocorrelation. For Beijing's traffic network, the first 25 autocorrelation coefficients exceeded the two-standard-error bands. This implies a strong autocorrelation behind the scaling behavior curve. For Guangzhou's traffic network, only the first 5 autocorrelation coefficients exceeded the two-standard-error bands. Compared with Beijing, the autocorrelation behind the scaling behavior curve of Guangzhou's traffic network is much weaker.)

### 3.4 Case analysis of four American cities

In the real world, fractal urban forms are complex and diverse. The fractal development stages and forms vary among different cities. City fractals are evolving structure, not existed structure from the beginning [36]. The scaling behavior curve can be used to identify the development status of cities and distinguish different types of fractal structures. Of course, if the fractal pattern has not developed to a convincing level, the behavior curve belongs to a non-scaling behavior curve. The fractal characteristics of urban growth and morphology can be examined from multiple different perspectives, including urban population, urban land use, urban nighttime light (NTL), and urban traffic network. Now, 7 European and American cities in 2 years are taken as examples to show the uses of scaling behavior curves based on NTL datasets. The 7 cities include New York, Chicago, Houston, Los Angeles, London, Paris, and Moscow. The two years are 2000 and 2022. The two years are 2000 and 2022. The data sources and quality were specifically explained in Ref. [37]. Due to the length of the paper, only the cases of New York, Chicago, Houston, and Los Angeles in 2000 are illustrated. The situation in other cities or other years can be analyzed in similar way (Supplementary Data S4).

A growing fractal follows power law, and correspondingly, the density decay curve of a growing fractal can be modeled by an inverse power function. As indicated above, Smeed's model was proposed to describe density decay of fractal cities [6]. However, the attenuation curves of NTL distribution density in these four American cities can be approximately described with Clark's model rather than the Smeed's model (Figure 10). Clark's model bears a parameter representing characteristic scale [12]. So, Clark's density decay used to be regarded as non-fractal distribution [6, 12]. However, the structure of the real world is diverse and complex, with true structure and false appearance coexisting. A discovery is that, if area- or number-radius scaling method is utilized to measure and calculate the dimension of a self-affine fractal, the density decay curve can be approximately modeled with a negative exponent function rather than an inverse power function [38]. The difference is as follows. The true exponential distribution can only be described by negative exponential functions and cannot be fitted by other functions. The pseudo exponential distribution caused by self-affine growth can also be modeled with two power functions in two segments. That is to say, self-affine fractals exhibit bi-fractal characteristics. The scale-measure relationship of bi-fractals takes on two sections of straight lines with different slopes on a log-log plot [34, 39].

Models are different from truth. Truth is unique, models are not unique. The same phenomenon can sometimes be described using different mathematical models. For example, there are many mathematical models for urban density [6, 40]. There is another situation, which can be treated as hidden fractals. The dual superposition of urban structure that satisfies Clark's model and the deep urban structure that satisfies Smeed's model will result in urban populations, land use, and NTL that satisfy the fractional gamma model (FGM) [40]. The standard form of the FGM can be expressed as

$$\rho(r) = Cr^{D-d} \exp(-\frac{r}{r_0}) , \tag{45}$$

where $C$ denotes the proportionality constant, $d$=2 refers to the Euclidean dimension of embedding space, and $D$ is the hidden fractal dimension. Taking the logarithm on both sides of equation (45) yields a linear equation. So the parameter values of equation (45) can be easily estimated by using multiple linear regression analysis based on the least squares calculation. If both global statistics (e.g., $R^2$, $F$ statistic) and local statistics (e.g., $t$ statistics or $P$ values) are significant and the parameter values are reasonable (e.g., $0<D<d$), it can be considered that urban density distribution meets FGM.

**Table 3 Parameter values and statistics of fitting FGM to density attenuation curves of four American cities (2000)**

| Parameter/statistics | New York | Chicago | Houston | Los Angeles |
|---|---|---|---|---|
| **Coefficient $C$** | 73.4576<br>($P$=0.0000) | 98.7030<br>($P$=0.0000) | 36.5566<br>($P$=0.0000) | 84.6389<br>($P$=0.0000) |
| **Characteristic radius $r_0$** | 23.6721<br>($P$=0.0000) | 40.0674<br>($P$=0.0000) | 14.4487<br>($P$=0.0000) | 37.8751<br>($P$=0.0000) |
| **Fractal dimension $D$** | 1.9974<br>($P$=0.9337*) | 1.8405<br>($P$=0.0000) | 2.2517**<br>($P$=0.0184) | 1.8584<br>($P$=0.0001) |

| **Goodness of fit $R^2$** | 0.9853 | 0.9788 | 0.9126 | 0.9708 |
|---|---|---|---|---|
| ***F* statistic** | 1878.0968 | 1406.1893 | 271.3509 | 1012.8950 |

**Note**: * This *P*-value implies a confidence level of 6.63%, which is very low.** When examining fractals in a two-dimensional embedding space, the fractal dimension value should not exceed 2. The value *D*=2.2517 is abnormal.

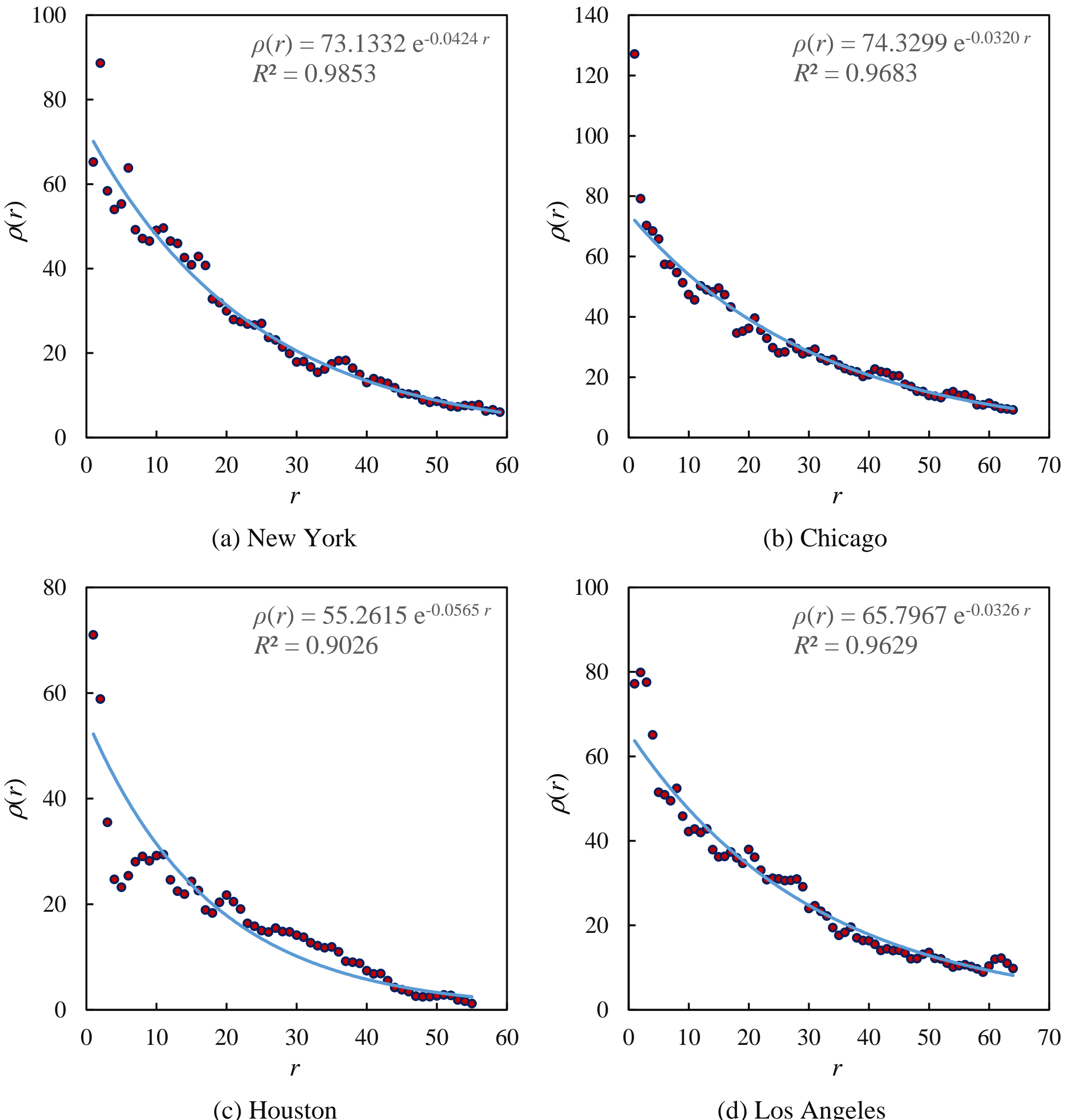

**Figure 10 The NTL density decay curves of New York, Chicago, Houston, and Los Angeles (2000)**

(**Note**: The NTL density decay curves of the four cities can be roughly described with negative exponential function. Among the four cities, the NTL density decay curves of Chicago and Los Angeles can also be described with FGM. The curve of New York cannot be described with FGM. The curve of Houston can be modeled with fractional gamma function, but the key parameter value cannot be reasonably explained.)

The results show that the density decay processes of Chicago and Los Angeles can be welled described with FGM, but the density decay curves of New York and Houston cannot be modeled with the fractional gamma function. For New York, the significance level of fractal parameter is as high as 0.9337, with a corresponding confidence level of less than 7%. This is statistically unacceptable. For Houston, based on a 95% confidence level, all model parameter values are statistically acceptable. But the fractal dimension exceeds the Euclidean dimension of the embedding space, so it is logically unacceptable (Table 3). For example, the FGM of Chicago in 2000 can be expressed as

$$\rho(r)=98.7030r^{1.8405-d}\exp(-\frac{r}{40.0674}), \tag{46}$$

which is both statistically and logically acceptable. The goodness of fit about $R^2$=0.9788, and the hidden fractal dimension is about $D$=1.8405. The fractal parameter value comes between 0 and 2 and meets theoretical expectations.

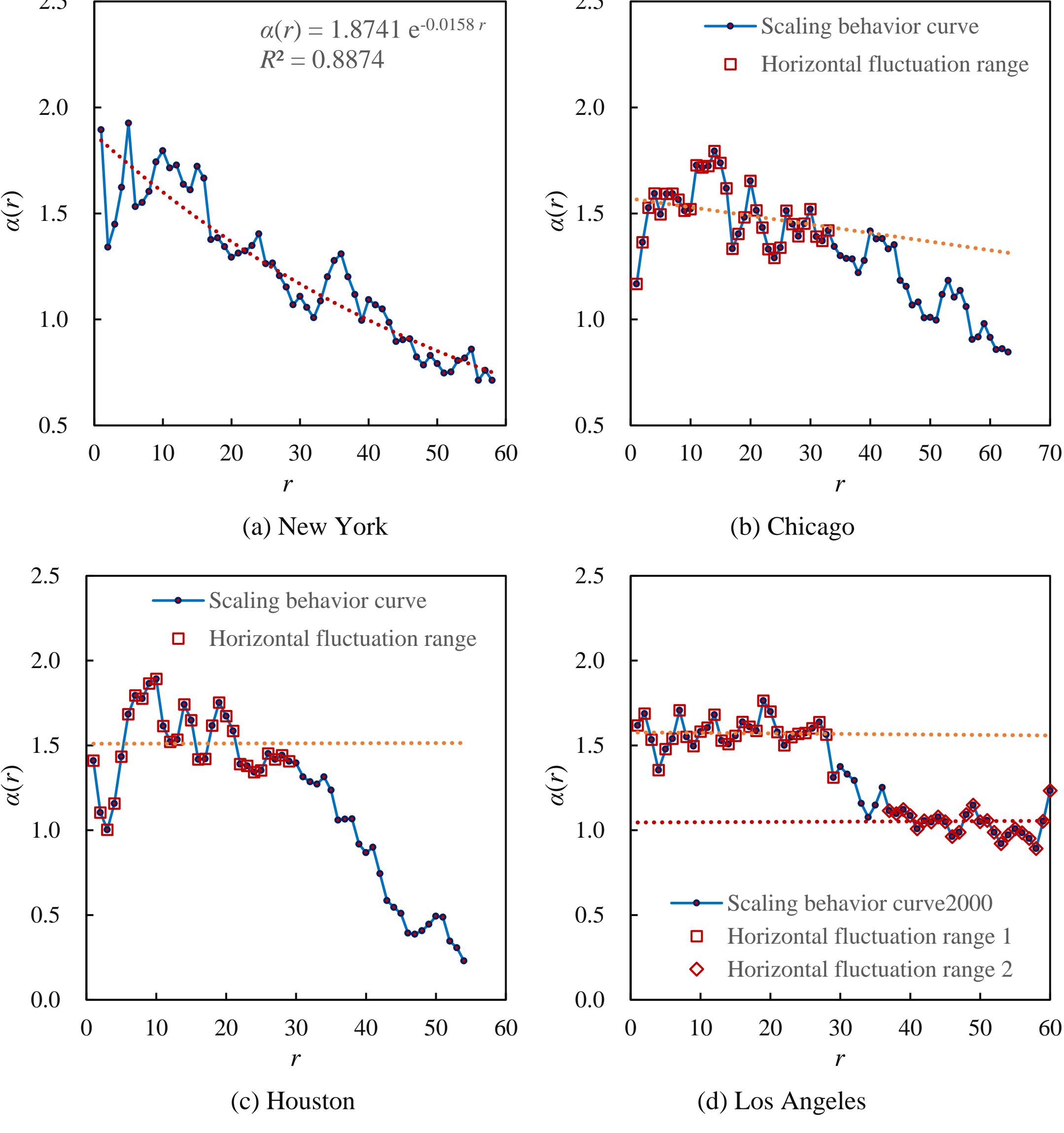

**Figure 11 The scaling/non-scaling behavior curves of NTL distributions of New York, Chicago, Houston, and Los Angeles (2000)**

(**Note**: Square hollow points represent the first horizontal fluctuation range, while diamond hollow points represent the second horizontal fluctuation range. The first and second subgraphs do not horizontal fluctuation range. The third subgraph has one horizontal fluctuation range, and the fourth subgraph has two horizontal fluctuation ranges.)

With the help of density attenuation curves and their mathematical models, we are still unable to effectively distinguish between Chicago and Los Angeles. Because the density decay curve of Los Angeles can also be described by the fractional gamma function. In this case, the scaling/non-scaling behavior curve can comes into play (Figure 11). The behavior curve of New York takes on negative exponential decay and has no scaling invariance. It is a non-scaling behavior curve. The situation of London and Paris is similar to that of New York. The first half of Chicago's behavior curve seems to fluctuate horizontally, but it cannot yield a horizontal trend line. The trend line of Chicago's behavior curve takes on a unimodal shape, suggesting a gamma curve. This is consistent with the FGM reflected by equation (46). In this sense, the behavior curve of Chicago can be regarded as a hidden fractal scaling behavior curve. The first half of Houston's behavior curve fluctuates horizontally. Adding trend lines to points within the horizontal fluctuation range can result in a horizontal trend line, the slope of which is close to 0. This suggests a locally developed fractal pattern. The situation of Moscow is similar to that of Houston. The behavior curve of Los Angeles has two horizontal linear fluctuation segments. Two horizontal fluctuation segments can generate two horizontal trend lines, indicating a bi-fractal structure. The behavior curves of Houston and Los Angeles can be treated as two types of local scaling behavior curves.

Next, we will examine the ACF histograms of the above behavior curves to determine whether they are scaling behavior curves. The correlation functions of the four cities have similar shapes and exhibit strong autocorrelation (Figure 12). Based on these correlograms, it is impossible to distinguish the significant differences between the four cities. This reflects the limitations of autocorrelation analysis of behavior curves. If we further examine the PACF, we will find that New York, Chicago, and Houston have only the first-order partial autocorrelation, while Los Angeles has high-order partial autocorrelation. One of the marks of negative exponential distribution is first-order partial autocorrelation. Higher order partial autocorrelation is a characteristic of power-law distributions. This seems to indicate that the urban form of Los Angeles is of bi-fractals, which can be described in two ranges using power functions with different fractal parameters. This also means that Los Angeles undergoes self-affine growth rather than self- similar growth.

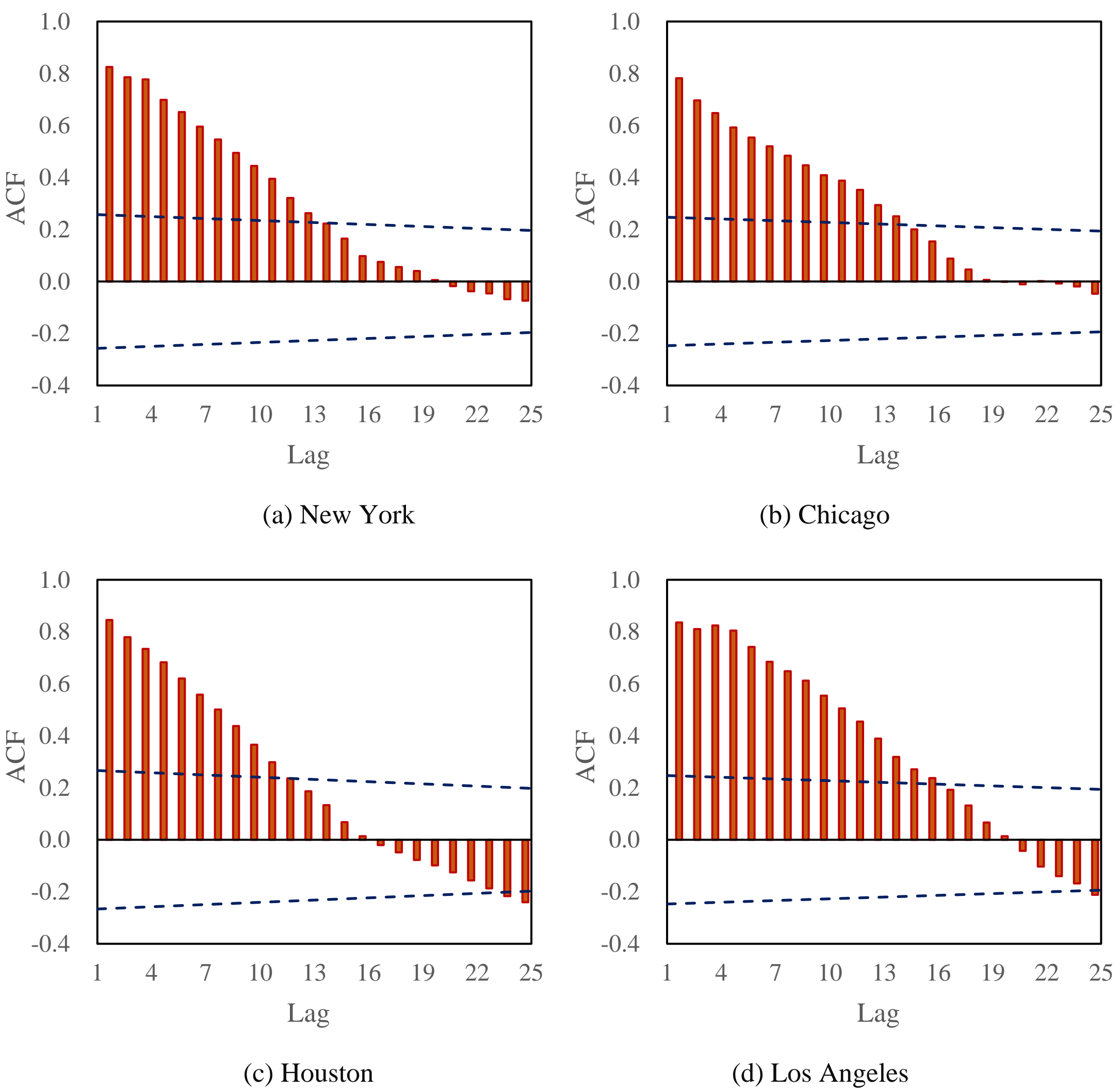

**Figure 12 Histograms of 1-dimensional spatial ACF of the scaling/non-scaling behavior curves of NTL distributions of New York, Chicago, Houston, and Los Angeles (2000)**

(**Note**: There are more than ten autocorrelation coefficients in each subgraph that exceed the two-standard-error bands, indicating strong autocorrelation.)

Finally, let's use the number-radius scaling method to examine the spatial distribution features and properties of the NTL in the four cities. If the entire urban cluster of a city exhibits a growing fractal, then equation (25) can be directly fitted to the NTL data. However, since the NTL density attenuation curve does not follow an inverse power law, it means that the growth fractal has not been fully developed in entire urban form. In this case, further judgment needs to be made using the scaling range. Two methods can be utilized to determine the scaling range. One is the *forward* *method*. Firstly, fit the fractal model, equation (25), to a few data points around the growth center. Then continuously increase the number of observation points outward until the goodness of fit reaches its maximum value. The other is *backward method*. Firstly, fit the fractal model, equation (25), to all the data points. Then continuously decrease the number of observation points inward until the goodness of fit reaches its peak value. The two methods have the same effect. The results are as follows (Figure 13). The scaling range of New York is very narrow. The number of points within the scaling range only accounts for 39% of all points. This is not enough to support the determination of fractal structure. Chicago and Houston have developed fractal structures within a certain range. The numbers of points within the scaling ranges account for 75% and 78% of all points, respectively. Los Angeles has two scaling ranges, indicating bi-fractal structure. This suggests that Los Angeles has anisotropic self-affine growth characteristics (Table 4).

**Table 4 NTL-based comparison of spatial characteristics of four American cities (2000)**

| **Item** | **New York** | **Chicago** | **Houston** | **Los Angeles** |
|---|---|---|---|---|
| **Density decay** | Negative exponential-like curve | Negative exponential-like curve | Negative exponential-like curve | Negative exponential-like curve |
| **Scaling/non-scaling behavior curve** | Non-scaling behavior curve | Hidden fractal scaling behavior curve | Local scaling behavior curve | Local scaling behavior curve |
| **ACF** | Multi order autocorrelation | Multi order autocorrelation | Multi order autocorrelation | Multi order autocorrelation |
| **PACF** | One order autocorrelation | One order autocorrelation | One order autocorrelation | Multi order autocorrelation |
| **Number of total data points** | 59 | 64 | 55 | 63 |
| **Number of data points within scaling range** | 23 | 48 | 43 | 29, 34 |
| **Proportion of scaling range points (%)** | 38.9831 | 75 | 78.1818 | 46.0317, 53.9683 |
| **Conclusion** | Non-fractal | Local fractal | Local fractal | Self-affine fractal |

In summary, based on NTL data, the following judgments can be made regarding the four cities. New York has not formed a convincing pattern of growing fractals. Chicago and Houston have growth fractal structure within a sufficiently large scale, but the growing fractal pattern does not encompass the entire urban form. Chicago exhibits a structure of fractal and non-fractal superposition, which is revealed by FGM. Los Angeles exhibits a bi-fractal structure, suggesting its self-affine growth characteristics. In addition, Los Angeles also embodies the overlapping characteristics of fractal and non-fractal. It should be emphasized that growing fractals are a type of fractal revealed through the radius area scaling relationship. The fractal properties of cities are multifaceted. Changing the measurement method or examining the perspective may lead to different conclusions.

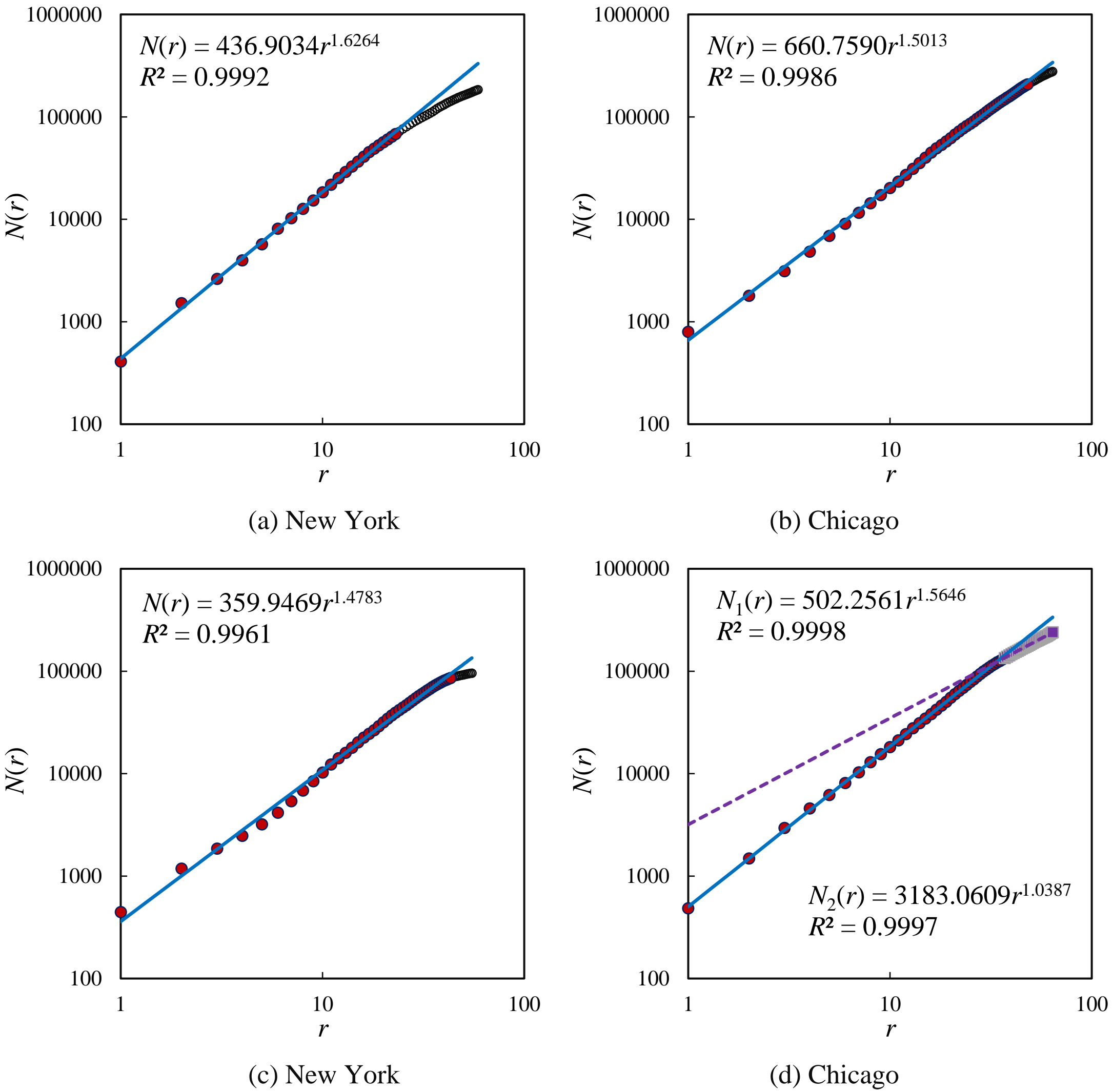

**Figure 13 The log-log plots for the relationships between the distance to city centers and total number of NTL of New York, Chicago, Houston, and Los Angeles (2000)**

(**Note**: The solid circular dots in each subgraph represent the first scaling range, the hollow circular dots represent the range outside the scaling range, and the solid square dots represent the second scaling range. Only Figure 13(d) has a second scaling range. As for the models, $r$ refers to the radius of concentric circles, and $N(r)$ to the total number of NTLs within the circle of radius $r$.)

## 4. Discussion

The formulae of fractal scaling behavior curves can be derived by three different methods. The first is to take the derivative of fractal dimension formulae, the second is to derive it from the model of self-similar hierarchies, and the third is to derive it from similarity dimension. Based on global fractal parameters and box-counting method, the fractal scaling behavior exponent is negative, while based on local fractal parameter and radius-area scaling method, the fractal scaling behavior exponent is positive. All these formula based on power law and inverse power law can be unified into the same logic framework. The basic property of a fractal scaling behavior curve can be reflected by ACF analysis. According to the fractal development state or fractal structure characteristics, the scaling behavior curves can be divided into three categories. Among these categories, the second category can be divided into two subclasses, and the third category can be divided into four subclasses (Table 5). If a city system is not fractal, the behavior curve bear no scaling nature and does not satisfy the equation (32). In this case, the behavior curve takes on a concave or convex curve. However, if the city system is of multifractal scaling, or self-affine scaling, the behavior curve will also be a concave or convex curve (Table 6). Self-affine fractal curve may be caused by two reasons: one is anisotropic growth, and the other is the wrong positioning of the growth center of isotropic growth.

**Table 5 Characteristics, criteria, and examples of three types of scaling behavior curves**

| Type | Characteristics | Criterion | Examples |
|---|---|---|---|
| **Global scaling** | The entire scaling behavior curve fluctuates randomly around a horizontal straight line | (1) No significant trend; (2) No significant autocorrelation: autocorrelation coefficients falls within two-standard-error bands | Figure 2(b); Figures 5(a) and 5(c) |
| **Local scaling** | The front or middle section of the scaling behavior curve fluctuates randomly around a horizontal straight line | (1) Trend is clear; (2) Weak autocorrelation: the first several autocorrelation coefficients exceed two-standard-error bands | Figure 5(b) and 5(d); Figure 8(b); Figure 11(c) and (d) |

| **Abnormal scaling** | The scaling behavior curve is manifested as an attenuation curve or a unimodal curve | (1) No significant trend; (2) Strong autocorrelation: many autocorrelation coefficients exceed two-standard-error bands | Figure 2(a); Figure 8(a); Figure 11(a) |
|---|---|---|---|

The curve of scaling behavior is a useful tool in the studies of fractal cities. First, the curves of scaling behavior can be used to evaluate the development level and property of fractal structure of cities. If a city bear fractal structure, both ACF and PACF come between the two standard error bands. In contrast, if the fractal structure of a city is not developed, partial values of ACF or PACF will break through the two standard error bands. On the other hand, if the fractal dimension depends on measurement scales, ACF or PACF will also go beyond the two standard error bands. In this case, the scaling behavior curve may reflect non-fractal phenomena, self-affine fractals, or multifractals. Second, the curves of scaling behavior can be utilized to identify the scaling range for fractal dimension estimation. In many cases, the power law relationships for fractal development only appear within certain scale limits. Reflected in log-log plots for fractal dimension estimation, a straight line segment appears in the scattered points. The straight line part is so-called scaling range. In empirical analyses, it is difficult to identify the lower limit and upper limit of the scaling range. Due to the sensitivity of scaling behavior curve to measurement scale, it can be employed to determine the scaling range of fractal development of cities. Third, the curves of scaling behavior can be employed to identify urban boundaries and growing center of fractal cities. For a fractal city, the lower limit of the scaling range indicates an equivalent circle of an urban envelope. An urban envelope is the closed urban boundary line [6, 41]. In this sense, the scaling behavior curve can be used to identify urban boundary [10]. In urban studies, it is significant to determine when and where a city fractal appears [36]. For radial fractal dimension, we have to identify properly the growing center of the city. In this respect, ACF analysis of fractal scaling behavior curve shows useful information for reveal urban growth center.

**Table 6 Three types of scaling behavior curves: Mathematical expressions and corresponding fractals**

| Type | Theoretical formula | Empirical formula | Scaling process |
|---|---|---|---|
| **Global scaling** | $\alpha_i = D$ | $\hat{\alpha}_i = D + \varepsilon_i$ | Monofractals |
| **Local scaling** | $\alpha_i = \begin{cases} d_E = 2, & r < r_l \\ D, & r_l \le r \le r_u \\ d_T = 0, & r > r_u \end{cases}$ | $\hat{\alpha}_i = \begin{cases} d_E + \varepsilon_i = 2 + \varepsilon_i, & r < r_l \\ D + \varepsilon_i, & r_l \le r \le r_u \\ d_T + \varepsilon_i = \varepsilon_i, & r > r_u \end{cases}$ | Monofractals |
| | | | Local developed fractal |
| **Abnormal scaling** | $\alpha_i = f(r_i)$ | $\hat{\alpha}_i = f(r_i) + \varepsilon_i$ | Scale-dependent fractal |
| | | | Self-affine scaling |
| | | | Multifractals |
| | | | Non-fractals |

The fractal scaling behavior exponent can be generalized from a constant to a function. For the local fractal model based on area-radius scaling, the derivative of fractal unit number $N(r)$ with respect to radius $r$ is

$$\alpha(r) = \frac{\mathrm{d}N(r)}{\mathrm{d}r} = DN_1 r^{D-1} = \frac{D}{r} N(r) . \tag{47}$$

Discretizing equation (47) yields

$$\alpha(r) = \frac{\mathrm{d}N(r)}{\mathrm{d}r} \to \alpha^*(r) = \frac{\Delta N(r)}{\Delta r} = \frac{N(r_i) - N(r_{i-1})}{r_i - r_{i-1}} . \tag{48}$$

It is easy to demonstrate that equation (46) follows scaling law, that is

$$\alpha(\zeta r) = \frac{D}{\zeta r} N(\zeta r) = \zeta^{D-1} \alpha(r) = \zeta^b \alpha(r) , \tag{49}$$

in which $\zeta$ denotes scale dilation factor, and the scaling exponent is

$$b = D - 1 , \tag{50}$$

where $D$ refers to the radial fractal dimension based on area-radius scaling. It is hard to make this form of scaling behavior curve clear in a few lines of words, and it requires a dedicated paper to discuss.

Further, the fractal scaling behavior curves analysis can be generalized to other fields involving scaling law. Scaling invariance is one of basic character of complex systems. Owing to scaling properties, a system bears no characteristic scale and thus cannot be described with conventional mathematical tools; owing to a system cannot be analyzed by conventional mathematical and quantitative methods, it is treated as one of complex systems. Besides fractals, famous scaling invariance phenomena include allometric growth, complex network, 1/*f* noise, and Zipf's law (Figure 14). Among these scaling invariance phenomena, Zipf's law and 1/*f* noise are the signatures of self-organized criticality and complexity [42]. Fractal geometry, allometry, and complex network theory compose the foundations for new science of cities [43, 44]. By introducing white noise into a mathematical model, we can simulate various scaling behavior curves for above-mentioned scale-free phenomena (Supplementary Data S5). All these scaling behavior curves show a horizontal trend and no significant autocorrelation (Table 7). The difference is that the scaling behavior exponent of allometric growth is positive, while the exponents of Zipf's distribution, 1/*f* noise and complex network are negative. The positive or negative value of the scaling behavior exponent depends on the form of the model describing the scaling process and does not reflect the essence of scaling invariance.

**Table 7 Similarities and differences between four types of scaling invariance phenomena**

| **Phenomenon** | **Model** | **Scaling behavior exponent** | **Trend** | **Autocorrelation** |
|---|---|---|---|---|
| **Allometric growth** | Positive power law | Positive | No trend | No autocorrelation |
| **Zipf's distribution** | Inverse power law | Negative | No trend | No autocorrelation |
| **1/*f* noise** | Inverse power law | Negative | No trend | No autocorrelation |
| **Complex network** | Inverse power law | Negative | No trend | No autocorrelation |

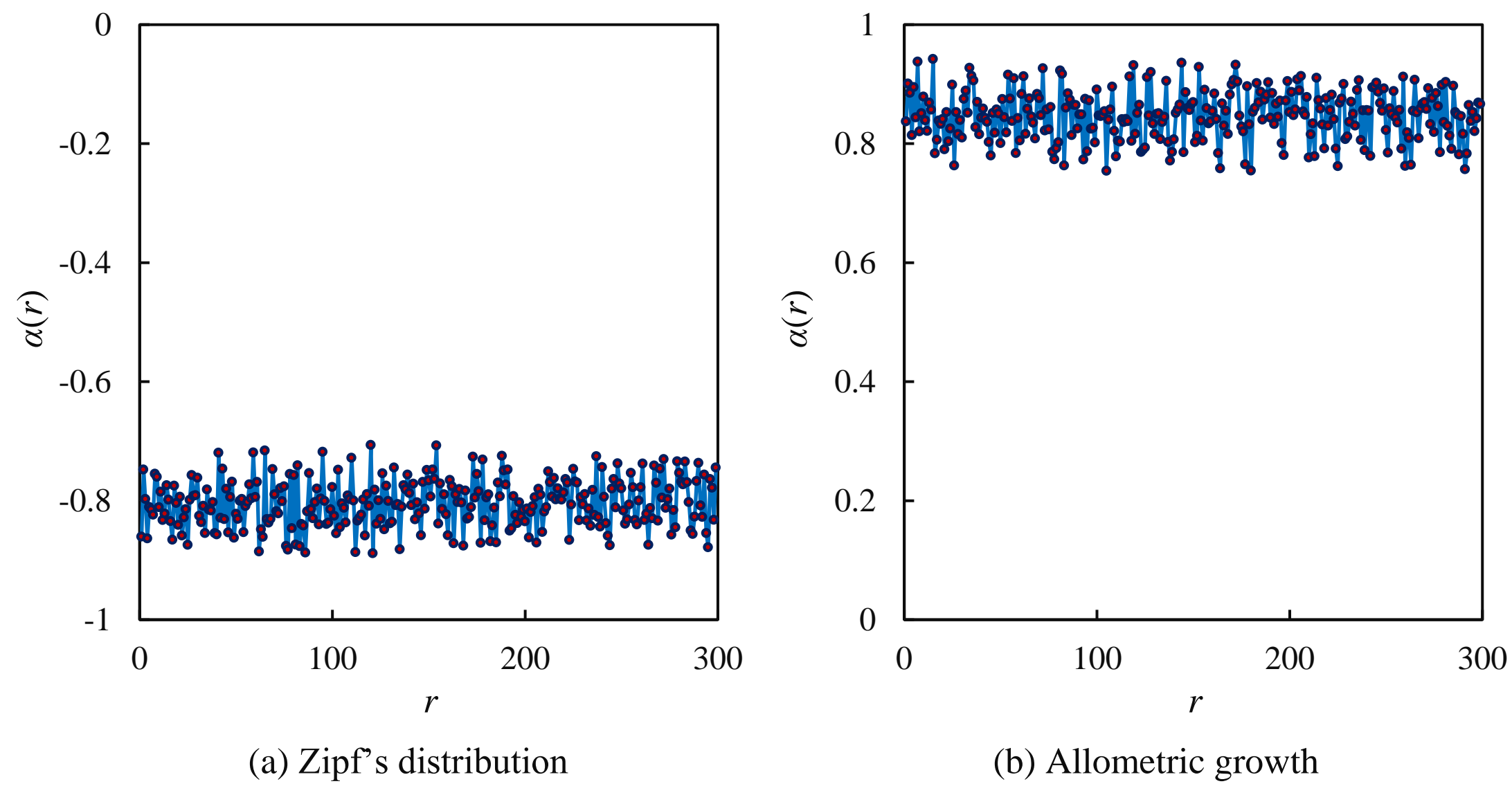

(a) Zipf's distribution (b) Allometric growth

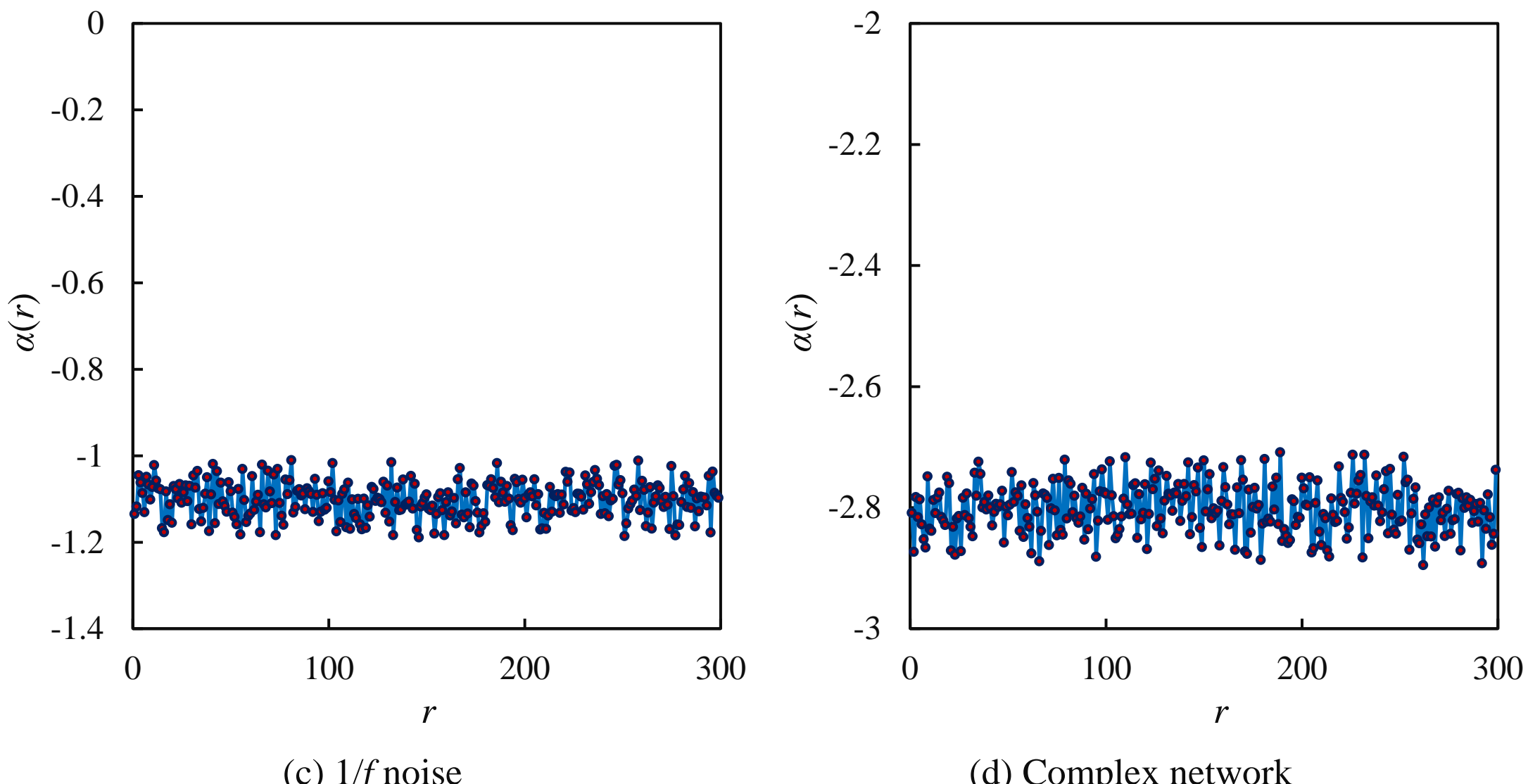

(c) 1/*f* noise (d) Complex network

**Figure 14 The scaling behavior curves of the general scaling phenomena: Zipf's law, allometric growth, 1/f noise, and complex network**

(**Note**: The scaling behavior curves were generated by means of Zipf's rank-size relation, allometric scaling law, 1/*f* frequency spectrum relation, and link-node relation in scale-free network of cities. See Supplementary Data S5. The commonality of these scaling behavior curves is that there is no significant trend. They fluctuate randomly around horizontal straight lines. This implies no autocorrelation behind the scaling processes.)

The novelty of this work rests with two aspects. First, a systematic logic framework of scaling behavior curves is presented. The formulae of scaling behavior curve are derived from different fractal models. Second, the autocorrelation and partial autocorrelation functions are utilized to analyze fractal scaling behavior. Correlograms and "two-standard-error bands" are employed to illustrate the statistical property of scaling behavior curves. The main shortcomings of this study is as below: First, the scaling behavior curve is based on monofractals rather multifractals. In the real world, the fractal systems are chiefly of multifractal structure instead of monofractal structure. Second, the four types of scaling behavior curves cannot be effectively distinguished, that is, non-fractals, self-affine fractals, multifractals, and scale-dependence fractals.

## 5. Conclusions

A fractal scaling behavior curve proved to be a curve of changing fractal dimension with measurement scales. Based on the theoretical derivation, numerical analysis, and empirical analysis, the main conclusions can be reached as follows. **First, a curve of fractal scaling behavior is a derivative or companion of a fractal model.** It can be derived from some kind of mathematical model of fractals. A fractal model is always based on cumulative distribution, while a fractal scaling behavior curve is based on density distribution. A density distribution is more sensitive to random disturbance or local differences than cumulative distribution. Therefore, the curves of fractal scaling behavior can be employed to reveal local feature or variety in details in fractal pattern or dynamic process. **Second, the curves of fractal scaling behavior can be divided into four types.** The first is the global scaling type. For regular fractals, this type of curve appears as a horizontal straight line; for random fractals, this type of curve fluctuates randomly and slightly around a horizontal straight line. The second is the local scaling type. The middle section of this curve is shown as a horizontal straight segment or an approximate horizontal straight segment. This type of curves reflects pre-fractals with local development. The third is the abnormal scaling behavior curve. No horizontal straight line or straight segment can be found or revealed. The scaling behavior of a possible fractal system shows a scale-dependence curve. This type of curves suggests self-affine fractals, multifractals, scale-dependence fractals, or non-fractals. **Third, 1-dimensional spatial ACF is an effective tools for revealing the statistic properties of scaling behavior curves of fractals**. In practice, ACF can be combined with PACF. For fractal structure, the ACF and PACF come within the "two-standard-error bands". If ACF and PACF go beyond the two-standard-error bands, the fractal structure is less developed, or it contain multifractal scaling process or self-affine scaling process or scale-dependence fractal process. Especially, ACF and PACF can be used to identify scaling range of fractal measurement. The upper limit of scaling range can be employed to find urban boundary line. For a growing city fractal, ACF and PACF can be used to identify the growth center of the city.

## Acknowledgement

This research was sponsored by the National Natural Science Foundation of China (Grant No. 42171192). The support is gratefully acknowledged.

# Data availability:

All data generated or analysed during this study are included in its supplementary materials.